\documentclass[12pt]{iopart}
\usepackage[pdftex]{graphicx}
\usepackage{amssymb}
\begin{document}

\title[FDTD solution of the Dirac equation and the Aharonov-Bohm effect]
{Finite Difference-Time Domain solution of the Dirac equation and the dynamics of the Aharonov-Bohm effect}

\author{Neven Simicevic \footnote[3]{Correspondence should be addressed to Louisiana Tech University, 
PO Box 10348, Ruston, LA 71272, Tel: +1.318.257.3591, Fax: +1.318.257.4228, 
E-mail: neven@phys.latech.edu}}

\address{\ Center for Applied Physics Studies, Louisiana Tech University,
 Ruston, LA 71272, USA}

\begin{abstract}

The time-dependent Dirac equation is solved using the three-dimensional
Finite Difference-Time Domain (FDTD) method. The dynamics of the electron wave packet 
in a vector potential is studied in the arrangements associated with the Aharonov-Bohm effect.
The solution of the Dirac equation showed a change in the velocity of the electron wave packet 
even in a region where no forces acted on the electron. The solution of the Dirac equation agreed
with the prediction of classical dynamics under the assumption that the dynamics was defined by 
the conservation of generalized or canonical momentum. It was also shown that in the case when 
the magnetic field was not zero, the conservation of generalized or canonical momentum was 
equivalent to the action of the Lorentz force.

\end{abstract}


\maketitle

\section{Introduction}

In our previous papers \cite{Sim08,Sim09}, the Finite Difference-Time Domain (FDTD) method, 
originally introduced by
Kane Yee \cite{Yee66} to solve Maxwell's equations, was for the first time 
applied to solve the three-dimensional Dirac equation. The $Zitterbewegung$ and 
the dynamics of a well-localized electron were used as examples of FDTD 
applied to the case of free electrons. The motion of electron wave packets 
inside and scattering from the potential step barrier or linearly dependent potential,
arrangements associated with the Klein paradox \cite{Klein29},  were
used as examples of interaction with a scalar potential. 
In this paper, a FDTD study of  the dynamics 
of an electron wave packet under the influence of a vector potential is presented. Such a dynamic behavior 
is most often associated with the Aharonov-Bohm effect \cite{AB59}. 

As a manifestation of quantum mechanics, charged particles passing around a long solenoid 
can feel a magnetic flux even when all the fields in the region through which the particles travel
are zero. The shifts in the phase of the wave functions describing the particles have been
experimentally verified by its effect on the interference fringes \cite {Cham60,Tono86,Osak86,Pesh89}.
Since, classically, there are no forces acting on the charged particles in the zero field region,
the theoretical description of the Aharonov-Bohm effect contains a number of assumptions. They include
assumptions on nonlocal features of quantum mechanics, the physical meaning of the vector potential,
topological effects, etc. Generally accepted physical understanding is still lacking \cite{Hege08}.
While there is still an open question on the presence of classical forces responsible for 
the Aharonov-Bohm effect \cite{Hege08,Boy06,Boy08}, on a macroscopic level they have 
not been observed \cite{Cap07}. 

Proper quantum-mechanical description of the dynamics of a relativistic charged particle involves 
the solution of the Dirac equation in the time domain. In addition to initial conditions, such a dynamics 
is defined only by the configuration of the scalar and vector potentials, and does not involve knowledge 
or any assumption on ``classical forces". The solutions of the time-dependent Dirac equation,
some of which are presented in this paper, can shed light and fill critical knowledge gaps on the theoretical
and experimental interpretations of the mechanism of the Aharonov-Bohm effect, including the existence
or non-existence of classical forces.

\section{Time-dependent solution of the Dirac equation}

The FDTD solutions of the time-dependent Dirac equation were obtained 
for the case when the
electromagnetic field described by the four-potential ${A^{\mu}=\{A_{0}(x),\vec A(x)\}}$
was minimally coupled to the particle \cite{Grein85,Sak87}
\begin{equation}
{\imath \hbar {\frac{\partial \Psi}{\partial t}}= ({H}_{free}+{H}_{int}) \Psi},
\label{Dirac_eq}
\end{equation}
where
\begin{equation}
{{H}_{free} = -\imath c\hbar {{\bf\alpha} \cdot \nabla} + \beta m c^{2}},
\end{equation}
\begin{equation}
{{H}_{int} = - e {{\bf\alpha} \cdot {\vec A} } + e A_{0}},
\end{equation}
and
\begin{equation}
{\Psi (x) =\left( \begin{array} {c} \Psi_{1} (x) \\ \Psi_{2} (x)
\\ \Psi_{3} (x)\\ \Psi_{4} (x) \end{array} \right)}.
\end{equation}
The matrices ${\bf\alpha}$ and $\beta$ were expressed using $2 \times 2$ Pauli
matrices $\bf\sigma^{'}s$ and the $2 \times 2$ unit matrix $I$.

In the FDTD method, the time dependent solution of the Dirac equation was obtained 
using updating difference equations \cite{Sim08}. As an example, the values of $\Psi_{1}$ at 
the position $(i\Delta x,j\Delta y,k\Delta z)$ and at the time step $(n+1/2)\Delta t$ were obtained 
using the equation

\begin{eqnarray}
 \Psi_{1}^{n+1/2}(I,J,K)&=&\frac{2-C^{n}(I,J,K)}{C^{n}(I,J,K)}\Psi_{1}^{n-1/2}(I,J,K) \nonumber \\
&-&{\frac{c\Delta t}{2\Delta x C^{n}(I,J,K)}}
[ \Psi_{3}^{n}(I,J,K+1)-\Psi_{3}^{n}(I,J,K-1) \nonumber \\
&+&\Psi_{4}^{n}(I+1,J,K)-\Psi_{4}^{n}(I-1,J,K)-i(\Psi_{4}^{n}(I,J+1,K) \nonumber \\
&-&\Psi_{4}^{n}(I,J-1,K))]
+i{\frac{e\Delta t}{\hbar C^{n}(I,J,K)}}[ A^{n}_{1}(I,J,K)\Psi_{4}^{n}(I,J,K) \nonumber \\
&-&iA^{n}_{2}(I,J,K)\Psi_{4}^{n}(I,J,K)
+A^{n}_{3}(I,J,K)\Psi_{3}^{n}(I,J,K)],
\label{Psi_1}
\end{eqnarray}
where $C^{n}(I,J,K)=1+i\frac{\Delta t}{2\hbar}[mc^{2}+eA^{n}_{0}(I,J,K)]$.
The space and time were discretized using uniform rectangular lattices of size
$\Delta x$, $\Delta y$ and $\Delta z$, and uniform time increment $\Delta t$.
While it is not generally required, in the Eq. (\ref{Psi_1}) $\Delta x = \Delta y = \Delta z$. 
Updating equations for $\Psi_{2}$, $\Psi_{3}$, and $\Psi_{4}$ were
constructed in a similar way.

The dynamics of a Dirac electron can now be studied
in any environment described by a four-potential $A^{\mu}$ regardless
of its complexity and time dependency. In this paper we studied the dynamics defined
by the vector potential $\vec A(x) \neq 0$.

The Dirac equation is a differential equation of the first order and linear
in $\partial / \partial t$. As in the case of Maxwell's equations, the entire 
dynamics of the electron is defined, only by its initial wave function. The dynamics
of a wave packet used in this paper was defined by the initial wave function of the form
\begin{equation}
{\Psi (\vec x,0) =N \sqrt{\frac{E+mc^{2}}{2E}}\left( \begin{array} {c} 1 \\ 0
\\ \frac{p_{3}c}{E+mc^{2}}\\ \frac{(p_{1}+ip_{2})c}{E+mc^{2}}\end{array} \right)}
e^{-\frac{\vec x \cdot \vec x }{4x_{0}^{2}}+\frac{i\vec p \cdot \vec x}{\hbar}},
\label{Wave_packet}
\end{equation}
where $N=[(2\pi)^{3/2}x_{0}^{3}]^{-1/2}$ is a normalizing constant. Eq. (\ref{Wave_packet})
represents a wave packet whose initial probability distribution is of a normalized Gaussian
shape. Its size is defined by the constant $x_{0}$, its spin is pointed along the z-axis, and its
motion is defined by the values of the momenta $p_{1},p_{2}$, and $p_{3}$. Some consequences of the 
initial localization of the wave packet on the overall dynamics of the electron were studied 
in Ref. \cite{Sim08,Huang52}.

\section{Validation of the computation: the dynamics of a wave packet in a strong uniform magnetic field}

The dynamics of a wave packet is very complex. The dynamics of a particle described 
by the wave packet in Eq. (\ref{Wave_packet}) depends on its localization, defined by 
the Gaussian component of the wave function, and its initial momentum, 
part of the wave function's phase. While an extensive study was done on the dynamics of
the wave packet related to the scalar component $A_{0}(x)$ of a four-potential \cite{Sim09}, 
such a study does not validate the dynamics of the wave packet related to the 
vector component of a four-potential $\vec A(x)$.

Applying the FDTD method to study the dynamics of a wave packet in a vector potential
associated with a strong uniform magnetic field is not difficult. Classically, the dynamics consist of
uniform rotational motion. 
In the relativistic quantum-mechanical description, however, even such a simple dynamics 
can validate the FDTD computation only up to a certain level. 
As pointed out in Ref. \cite{Schl08} "the dynamics is particularly
rich and not adequately described by semiclassical approximations". In Ref. \cite{Demi08} 
it was demonstrated that in the presence of an external magnetic field the wave packet splits into two parts 
which rotate with different cyclotron frequencies, and after a few periods, the motion acquires irregular
character.  As a result of such a complex dynamics, when comparing the results obtained by
the FDTD method and the computation described in Ref. \cite{Schl08} and \cite{Demi08},
we could not expect more than a qualitative agreement. In addition to this qualitative agreement,
of equal importance to the validation of the computation is the consistency of the results obtained 
for different vector potential gauges.

In this paper, the motion of the wave packet described by Eq. (\ref{Wave_packet}) 
was studied in a uniform magnetic field oriented along the y-axis

\begin{equation}
\vec B=(0,B_{0},0).
\label{Uni_mag_field}
\end{equation}
For this field the corresponding vector potential in a rotationally invariant gauge was
\begin{equation}
\vec A={B_{0} \over 2} (-z,0,x).
\label{Uni_mag_potent}
\end{equation}
The dynamics of the wave packet was obtained by solving the Dirac equation for this vector potential.
 
\begin{figure}
\centering
{\scalebox{.7}{\includegraphics{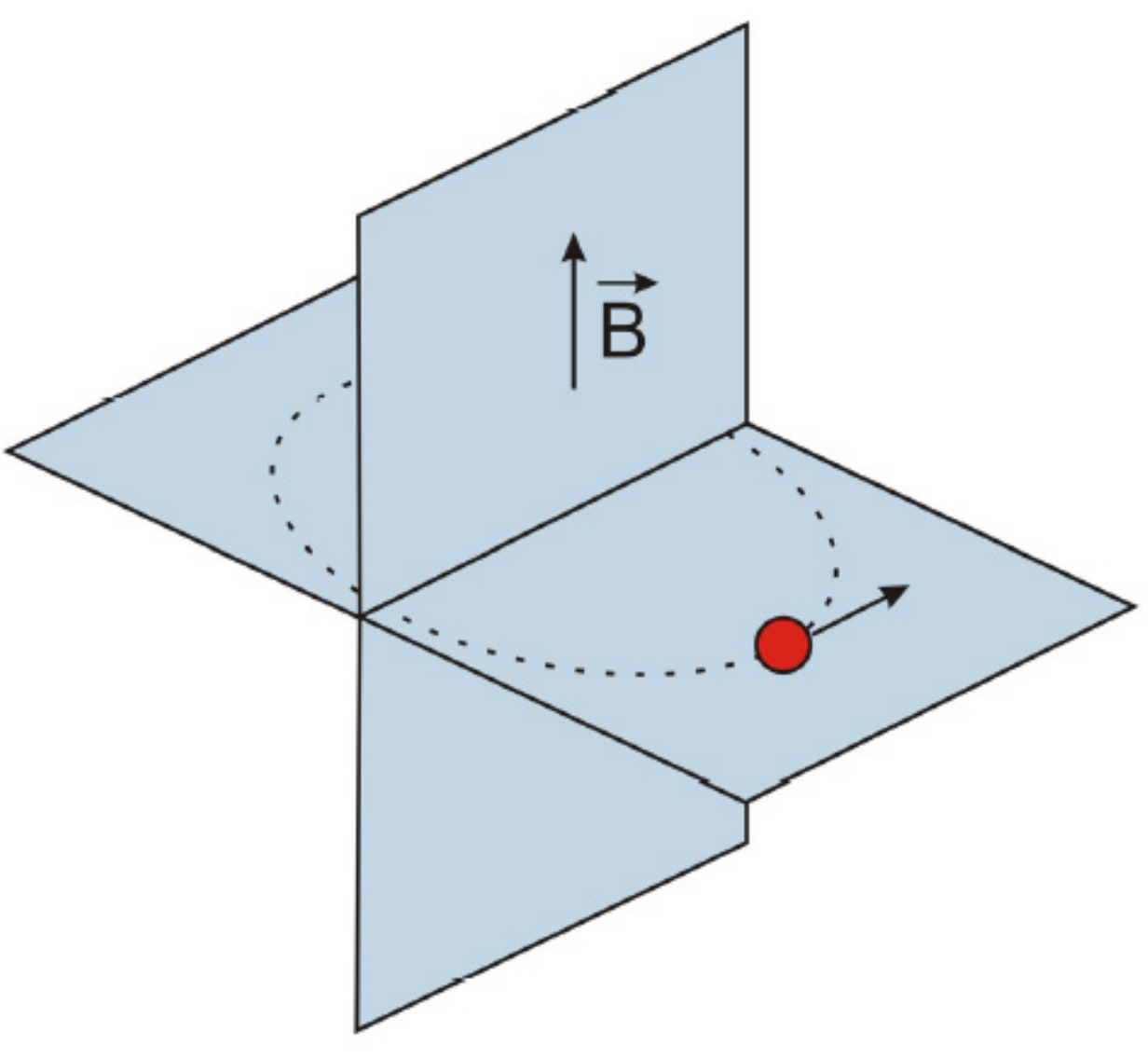}}}
\caption{\label{fig:Planes_unif_field} Schematics of the two planes used in displaying 
the probability densities. They are shown relative to the classical orbit of the electron
and the orientation of the uniform magnetic field.}
\end{figure}

\begin{figure}
\centering
\vspace*{- 3. cm}
{\scalebox{.45}{\includegraphics{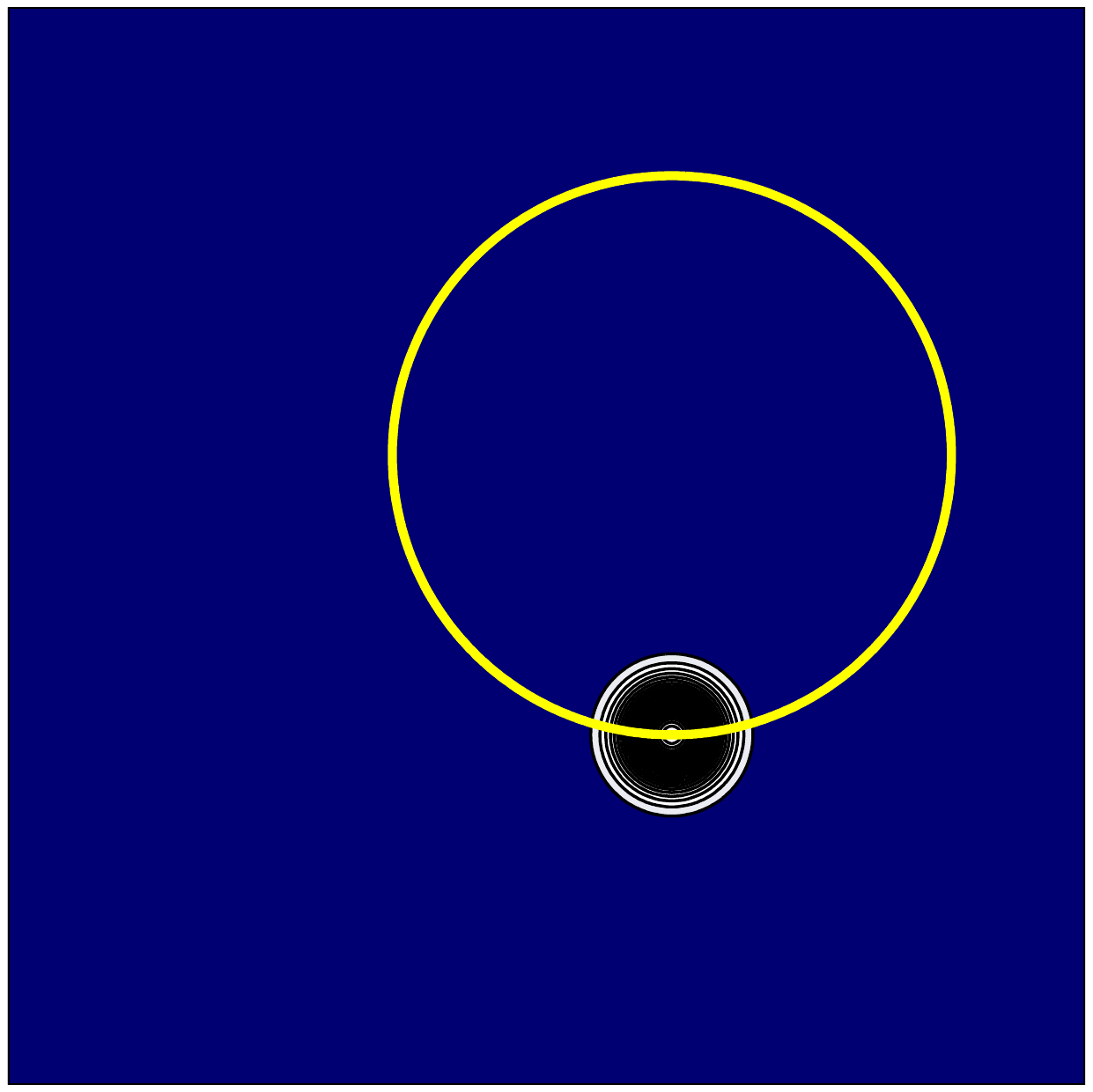}}}\hspace*{- 4.5 cm}
{\scalebox{.45}{\includegraphics{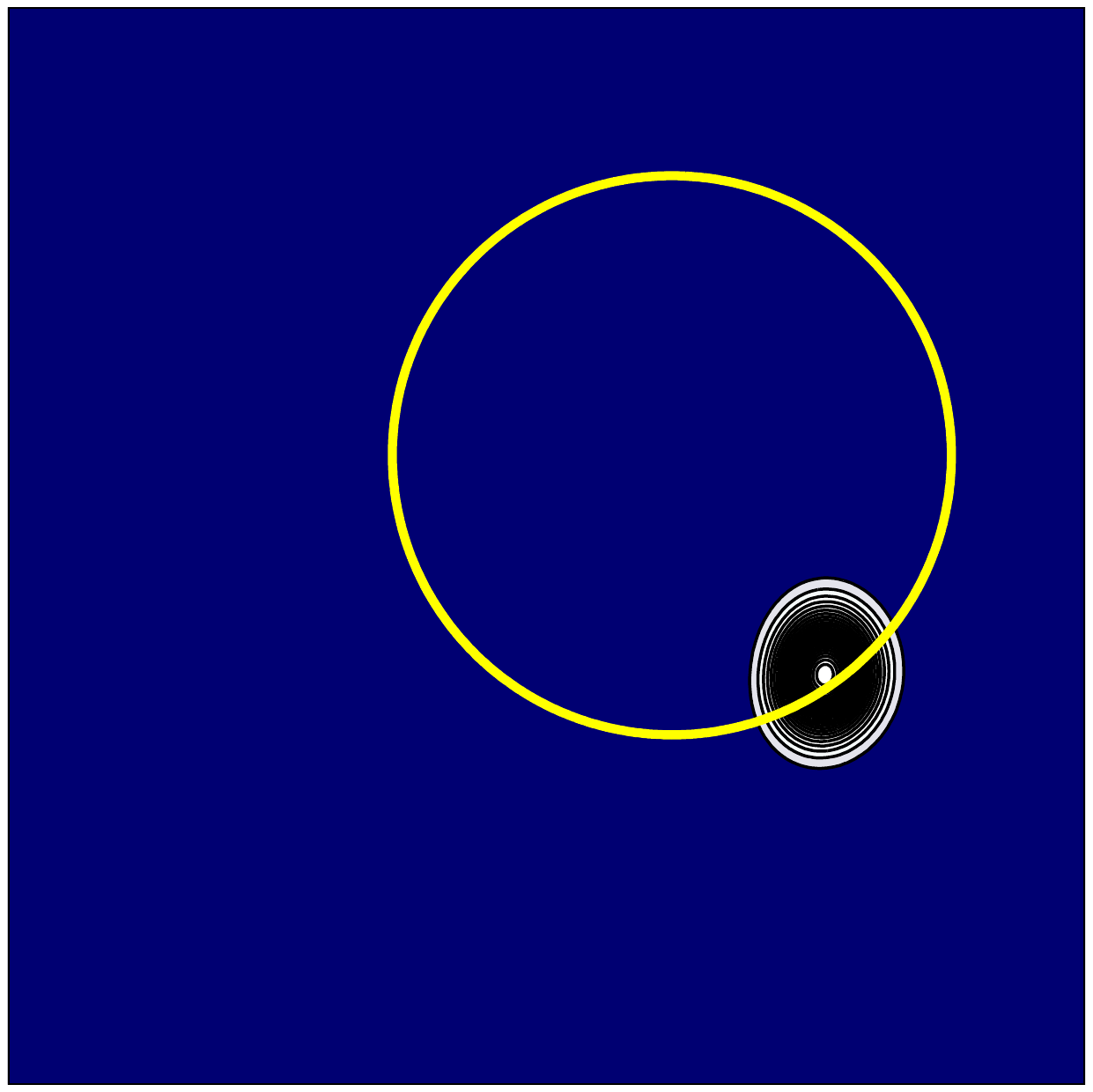}}}\vspace*{- 7.5 cm}
{\scalebox{.45}{\includegraphics{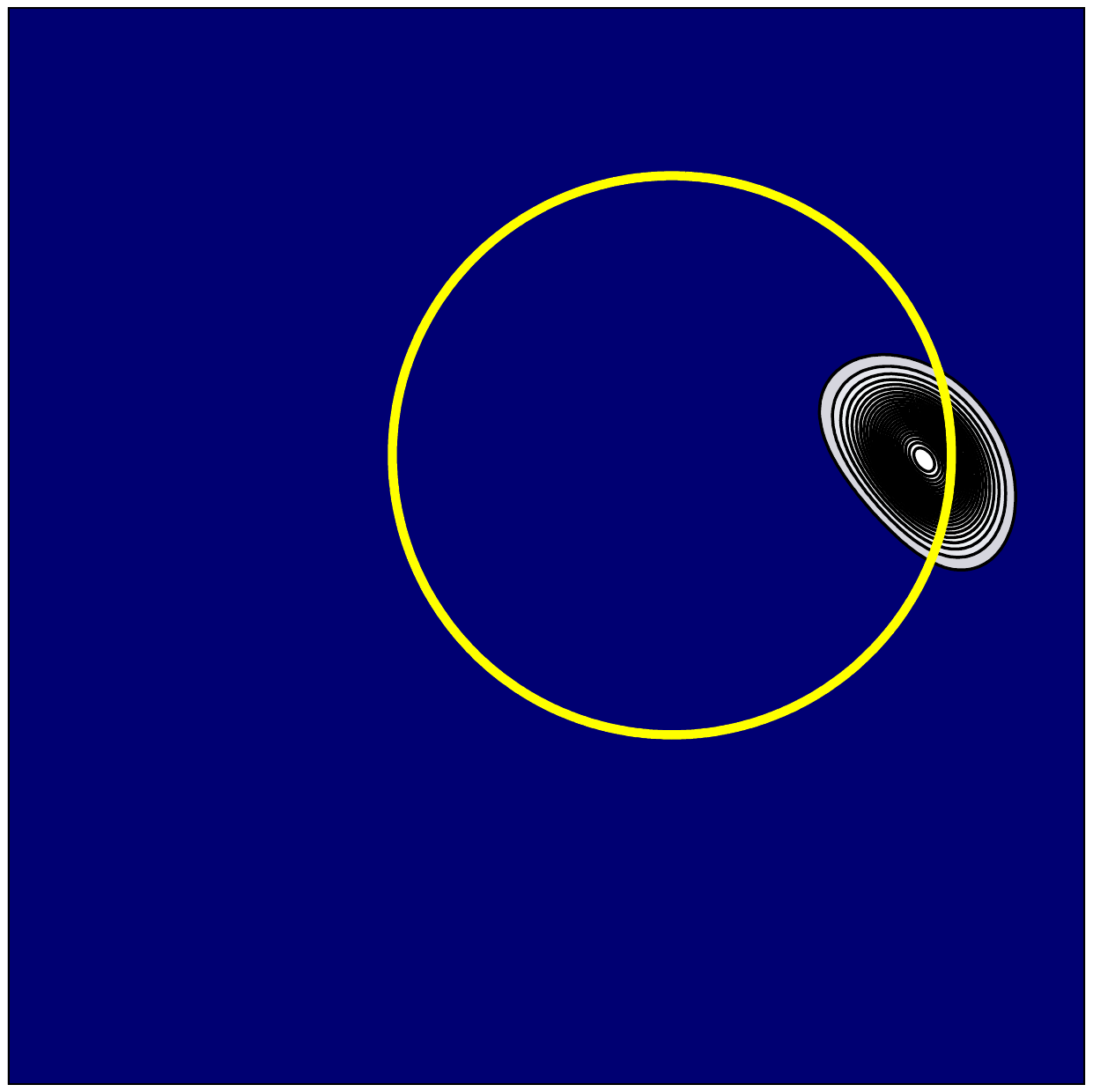}}}\hspace*{- 4.5cm}
{\scalebox{.45}{\includegraphics{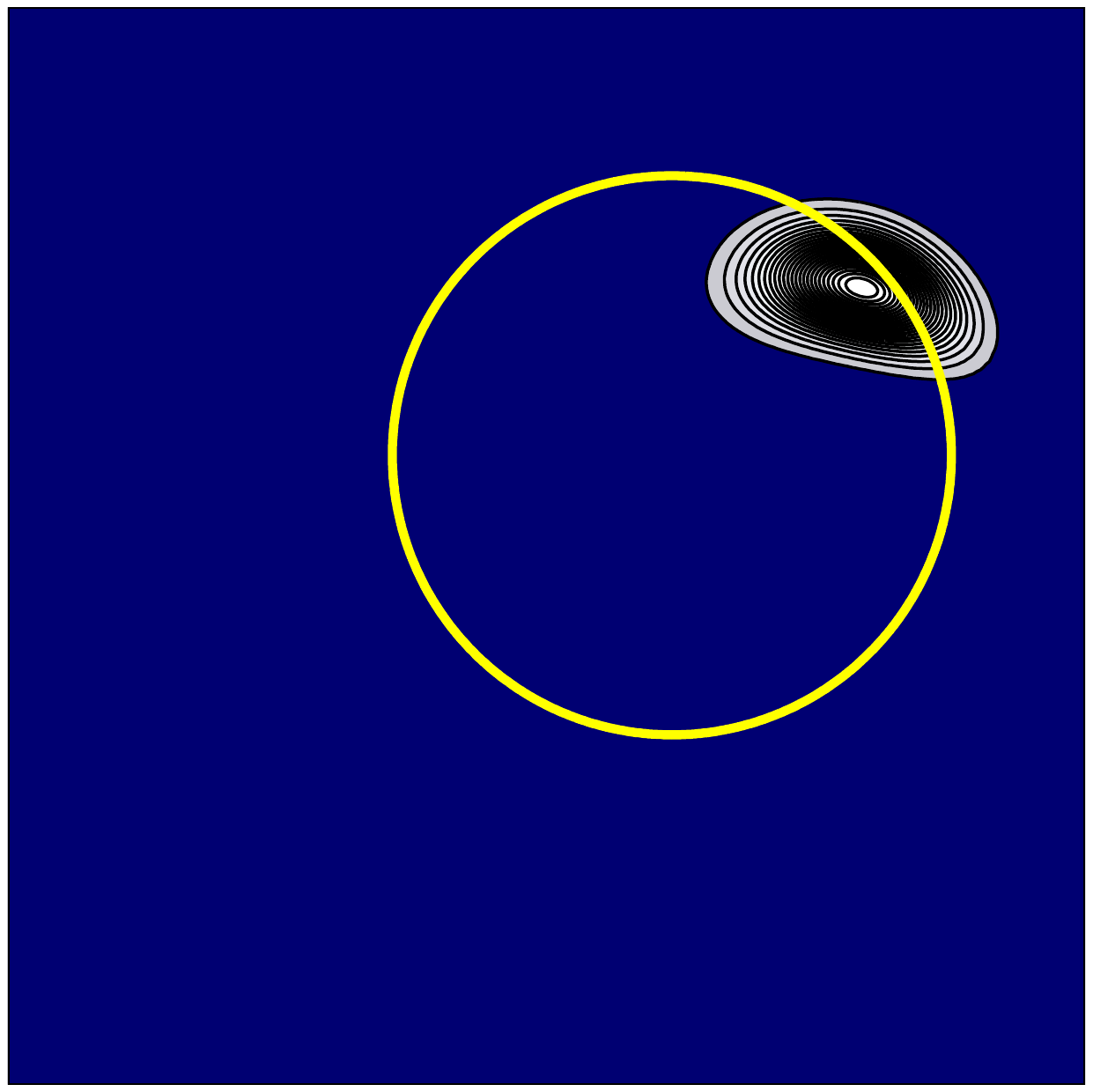}}}\vspace*{- 7.5 cm}
{\scalebox{.45}{\includegraphics{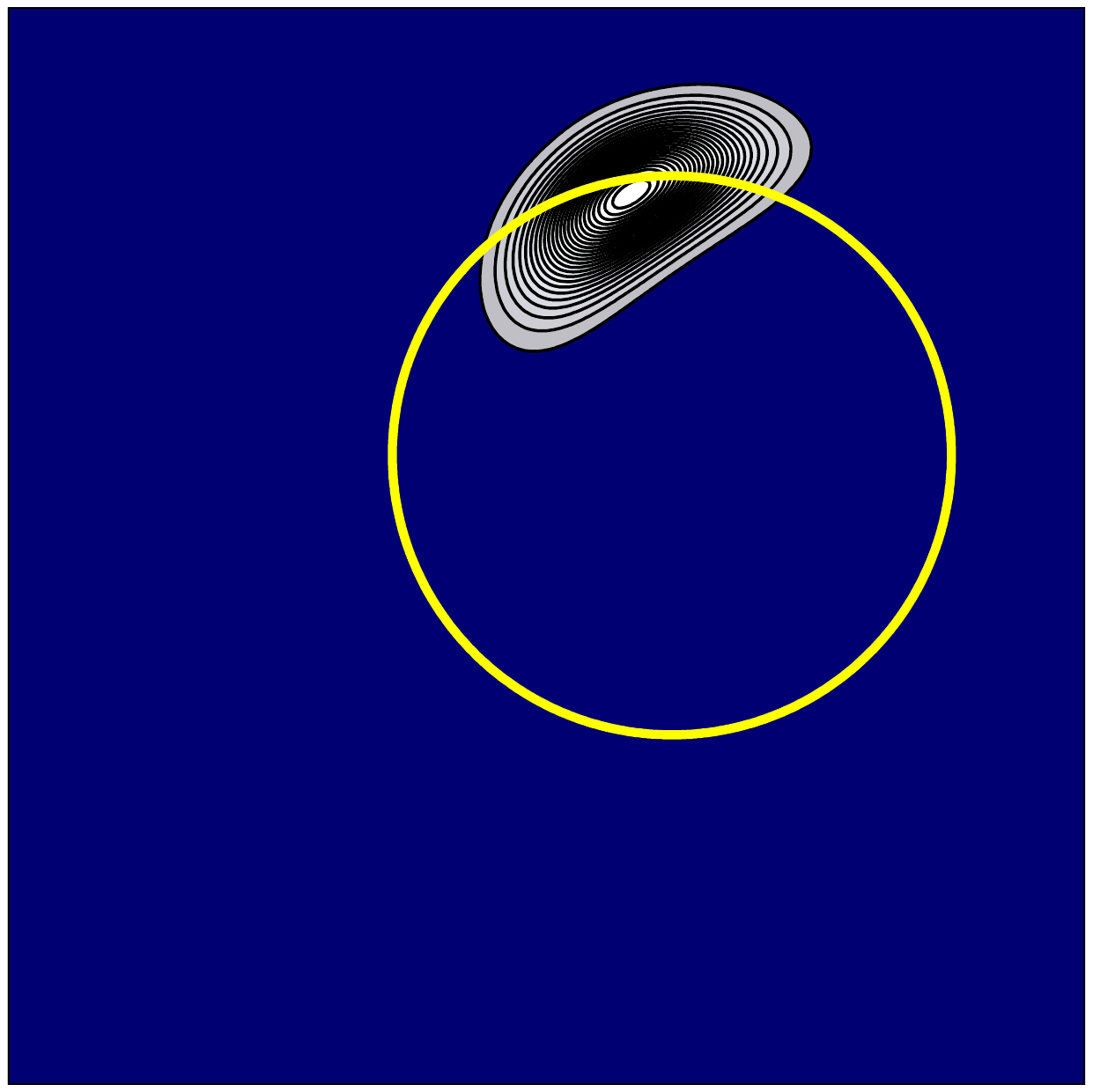}}}\hspace*{- 4.5 cm}
{\scalebox{.45}{\includegraphics{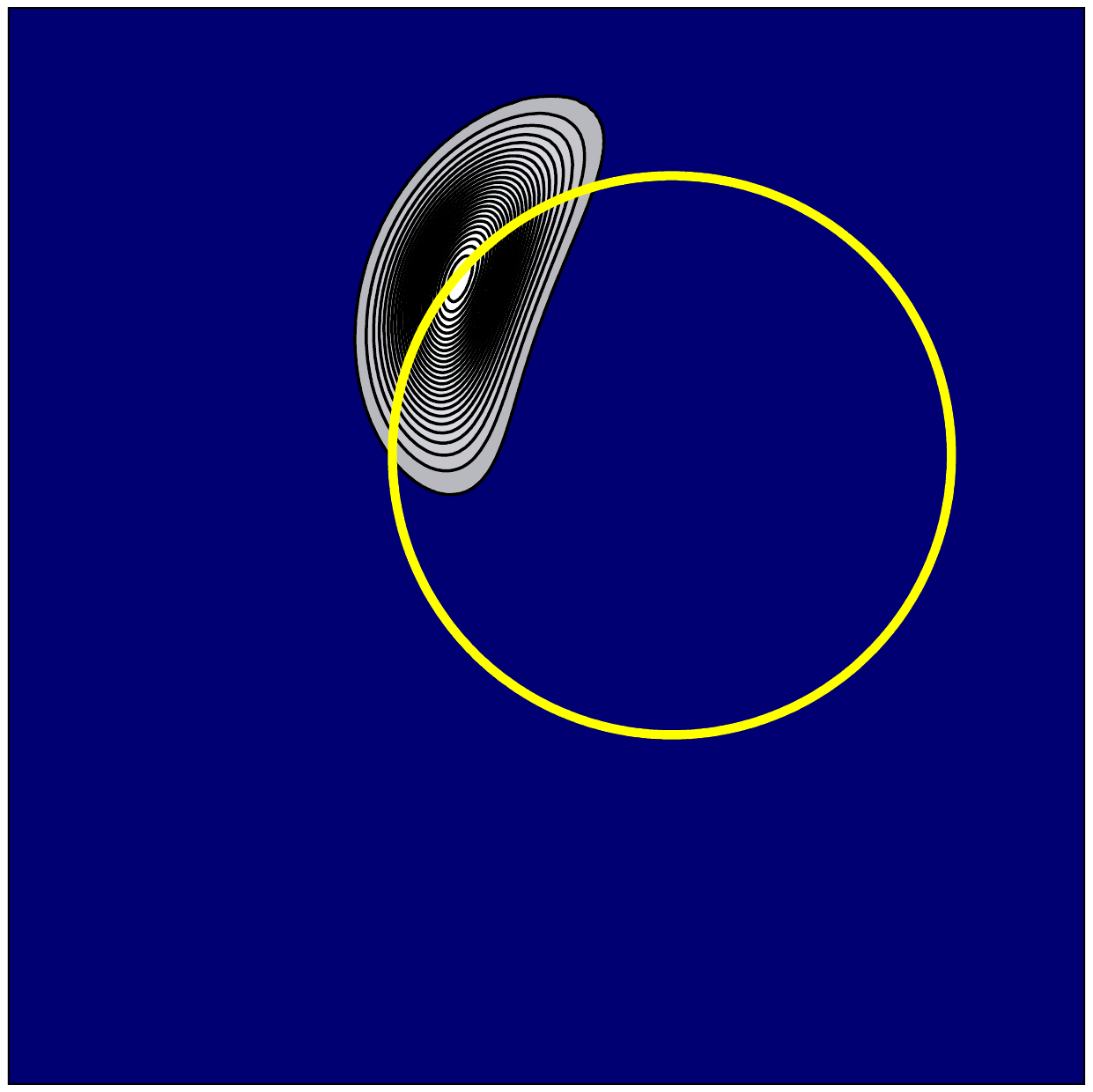}}}\vspace*{- 7.5 cm}
{\scalebox{.45}{\includegraphics{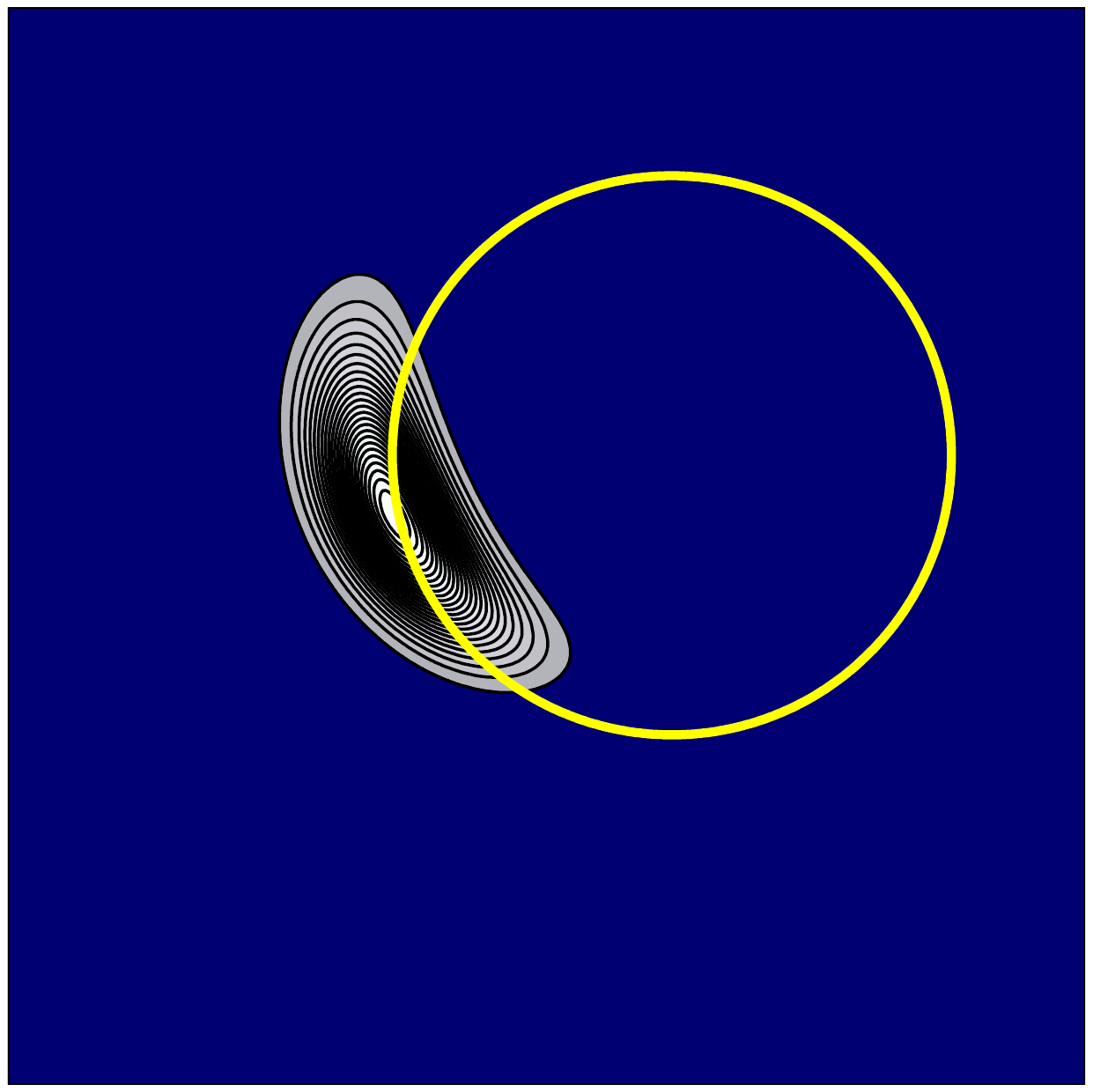}}}\hspace*{- 4.5 cm}
{\scalebox{.45}{\includegraphics{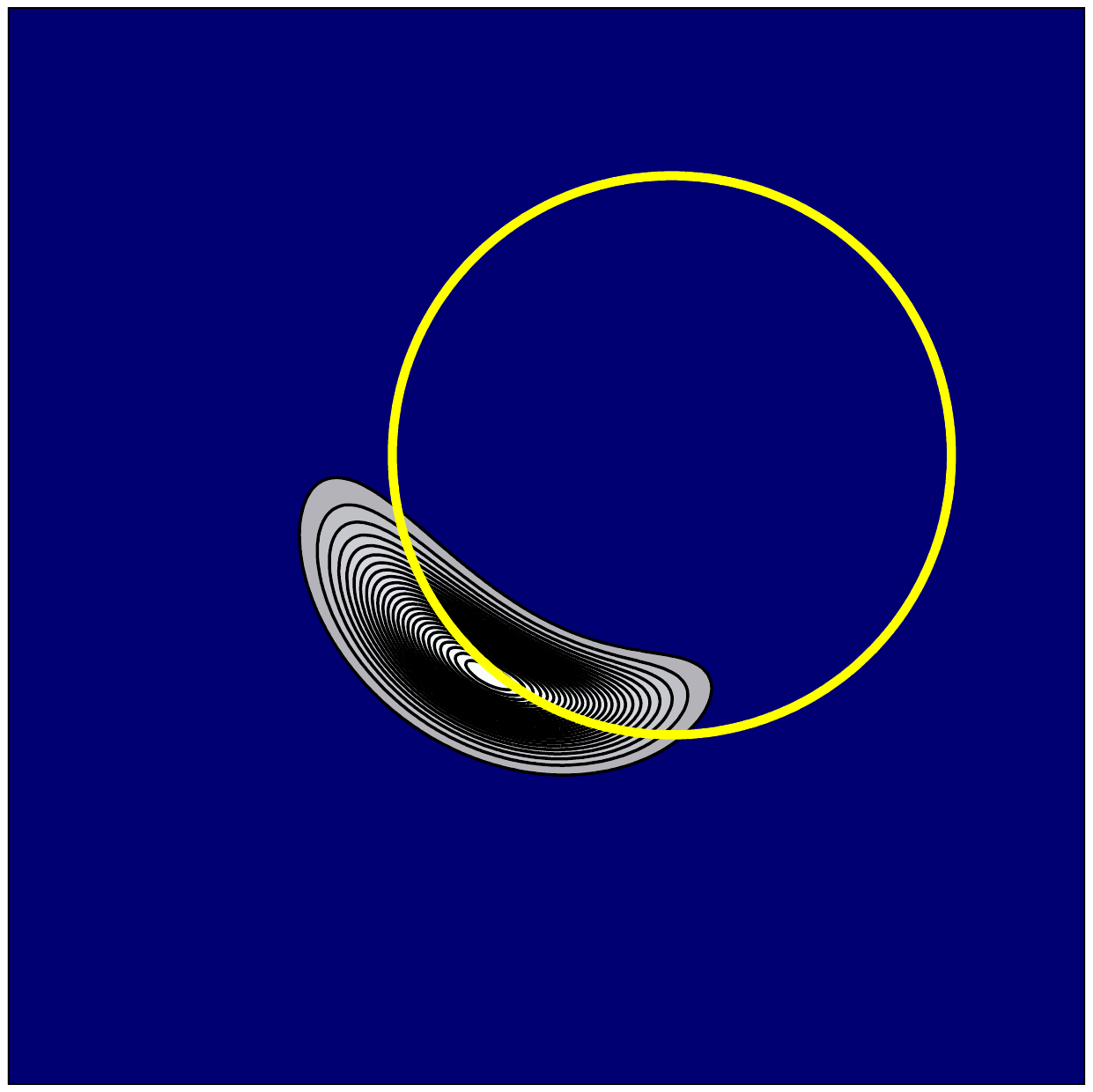}}}\vspace*{- 3. cm }
\caption{\label{fig:First_orbit_unif_field} The shapes and the positions of
the wave packet in the horizontal plane during the first rotation. The circle represents the classical orbit.
The animation can be accessed on-line \cite{Simi09b}.}
\end{figure}

\begin{figure}
\centering
\vspace*{- 4.5 cm}
{\scalebox{.5}{\includegraphics{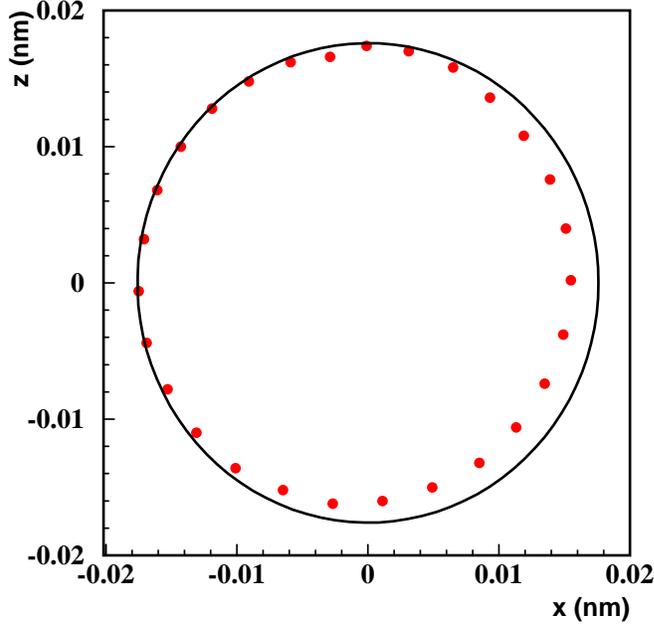}}}
\vspace*{- 1. cm}
\caption{\label{fig:First_orbit_position_cprob} Positions of the center of probability 
of the wave packet  in the horizontal plane during the first rotation. The circle represents the classical orbit.
The animation of the dynamics can be accessed on-line \cite{Simi09b}.}
\end{figure}

While the probability densities $|\Psi|^2$ were calculated for the entire computational volume and
at every time step, their values are shown here only
on two planes, the horizontal or plane of classical particle motion, and the plane vertical to the plane of motion. 
The schematics of the planes relative to the wave packet motion and the orientation of the magnetic field 
are shown in Fig \ref{fig:Planes_unif_field}. 
The position and the shape of the wave packet in the horizontal plane, as it moves along its first orbit,
is shown in Figure \ref{fig:First_orbit_unif_field}. The initial position of the wave packet was at the
center of the vector potential. Its initial momentum was $p_{1} = 0.53 \; MeV/c$, making the motion relativistic. 
In order to force a relativistic electron to complete a full circle in the available computational 
space, the magnitude of the magnetic field was $B_{0} = 10^{8} \; T$. The field of such 
a strength is associated with the fields at the surface of the neutron stars. The classical orbit of the electron in
this field is $r_{class}=p_{1}/(e B_{0})=1.76 \times 10^{-2}\; nm$. As shown in 
Figure \ref{fig:First_orbit_unif_field}, during the first rotation the wave packet 
generally follows the classical orbit and disperses at the same time. The position of the center of probability 
of the wave packet relative to the classical orbit is shown in Fig \ref{fig:First_orbit_position_cprob}. 

\begin{figure}
\centering
{\scalebox{.45}{\includegraphics{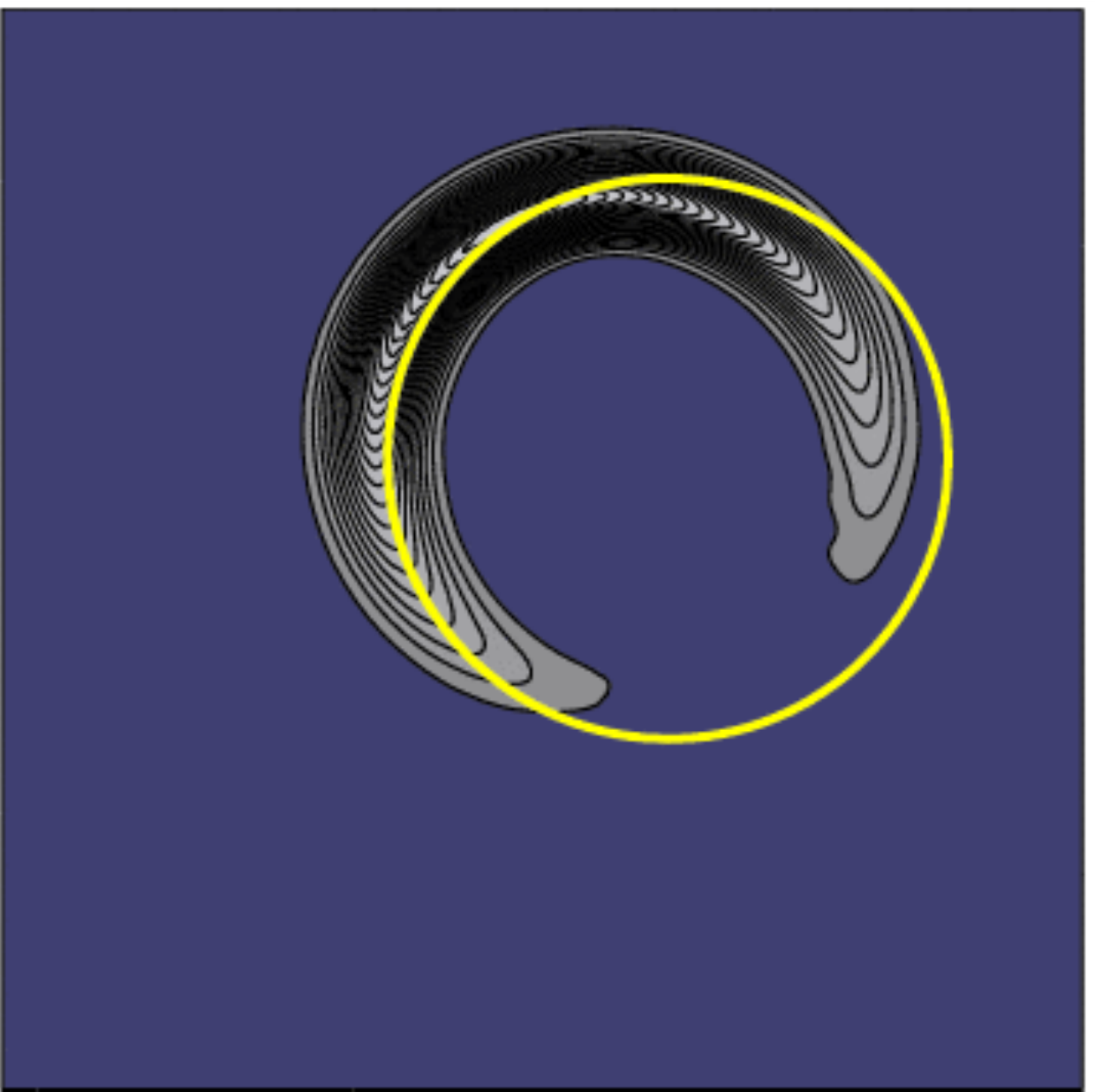}}}\hspace*{- 0.1 cm}
{\scalebox{.45}{\includegraphics{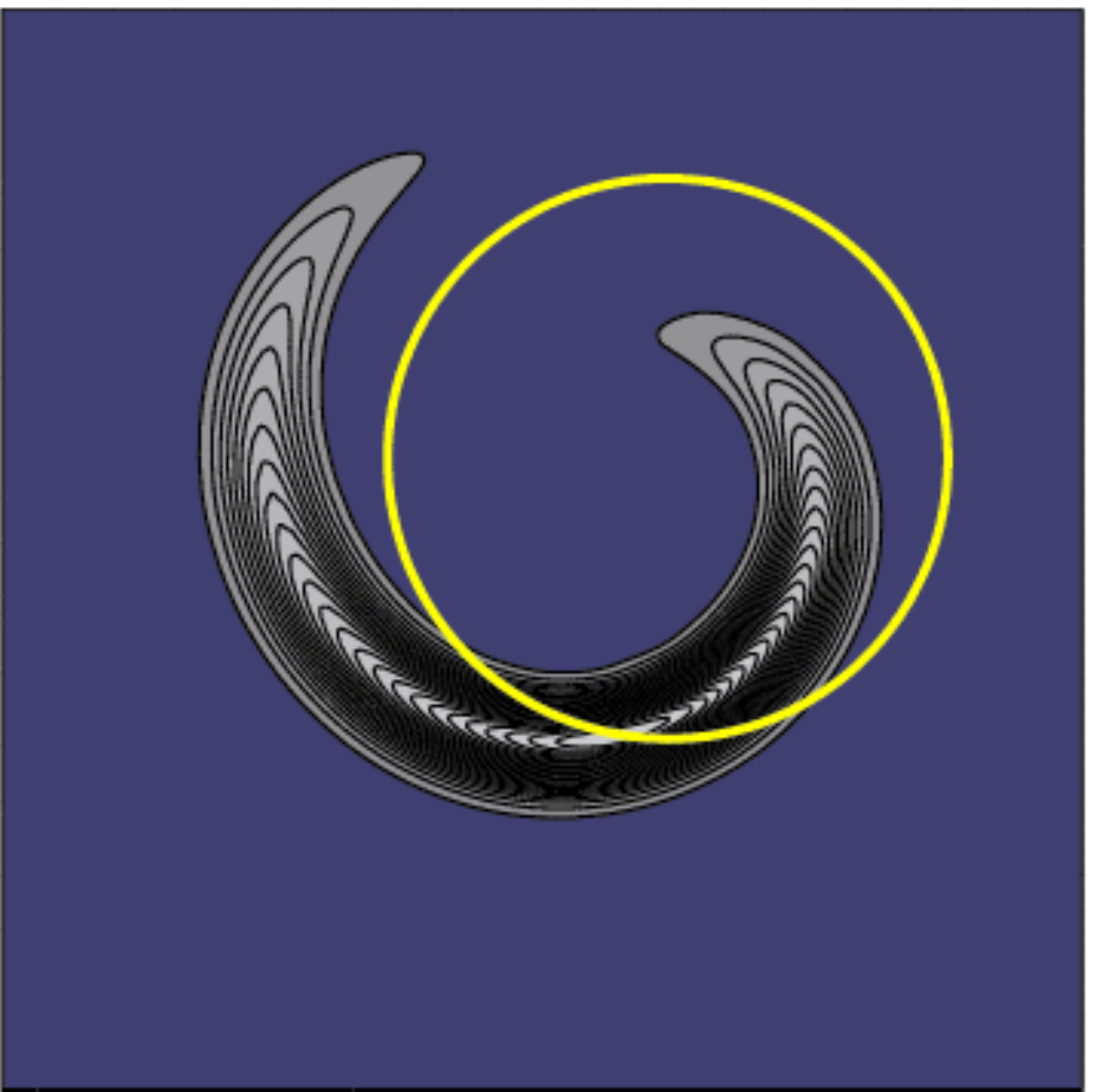}}}
{\scalebox{.45}{\includegraphics{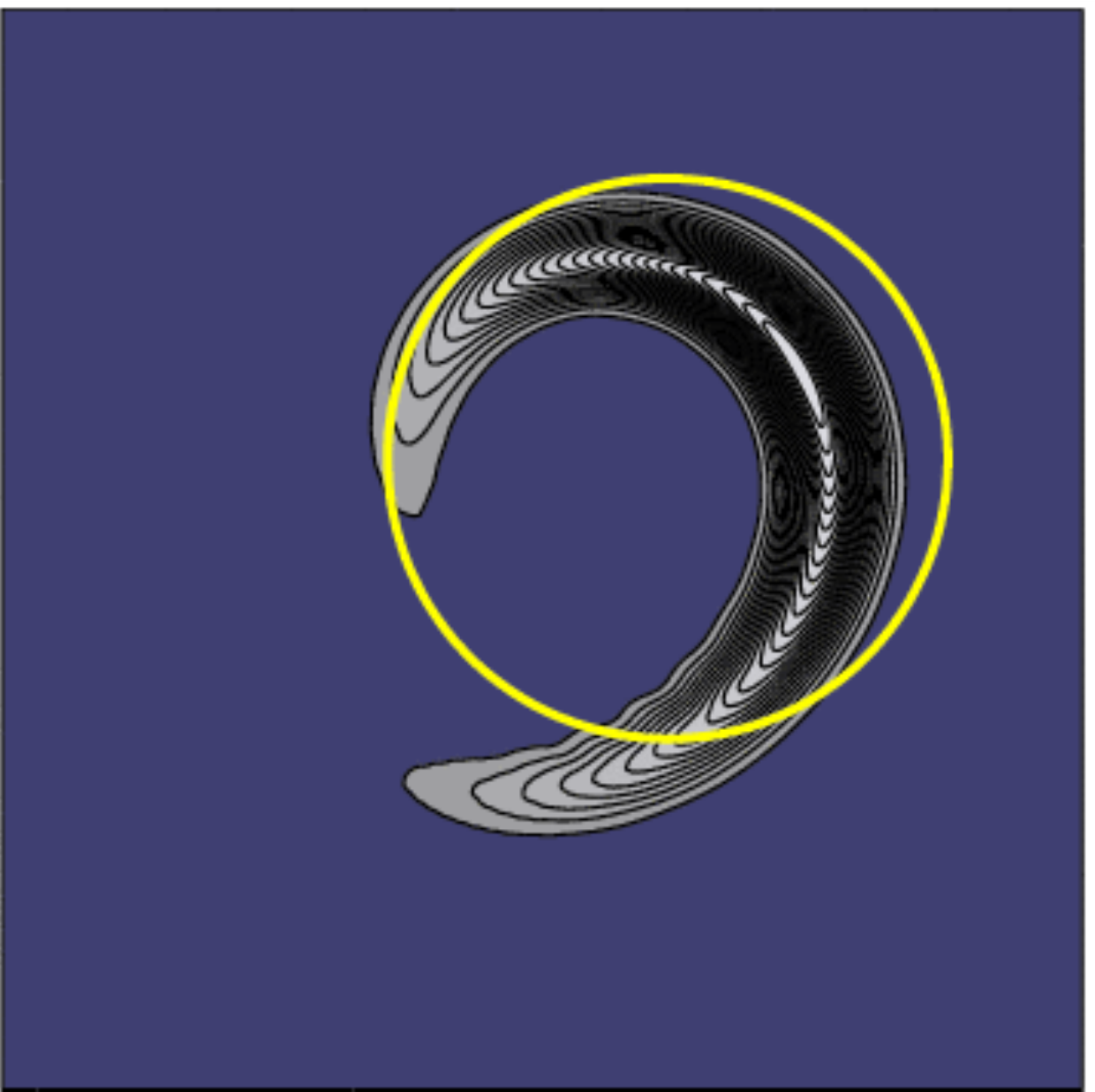}}}\hspace*{- 0.1 cm}
{\scalebox{.45}{\includegraphics{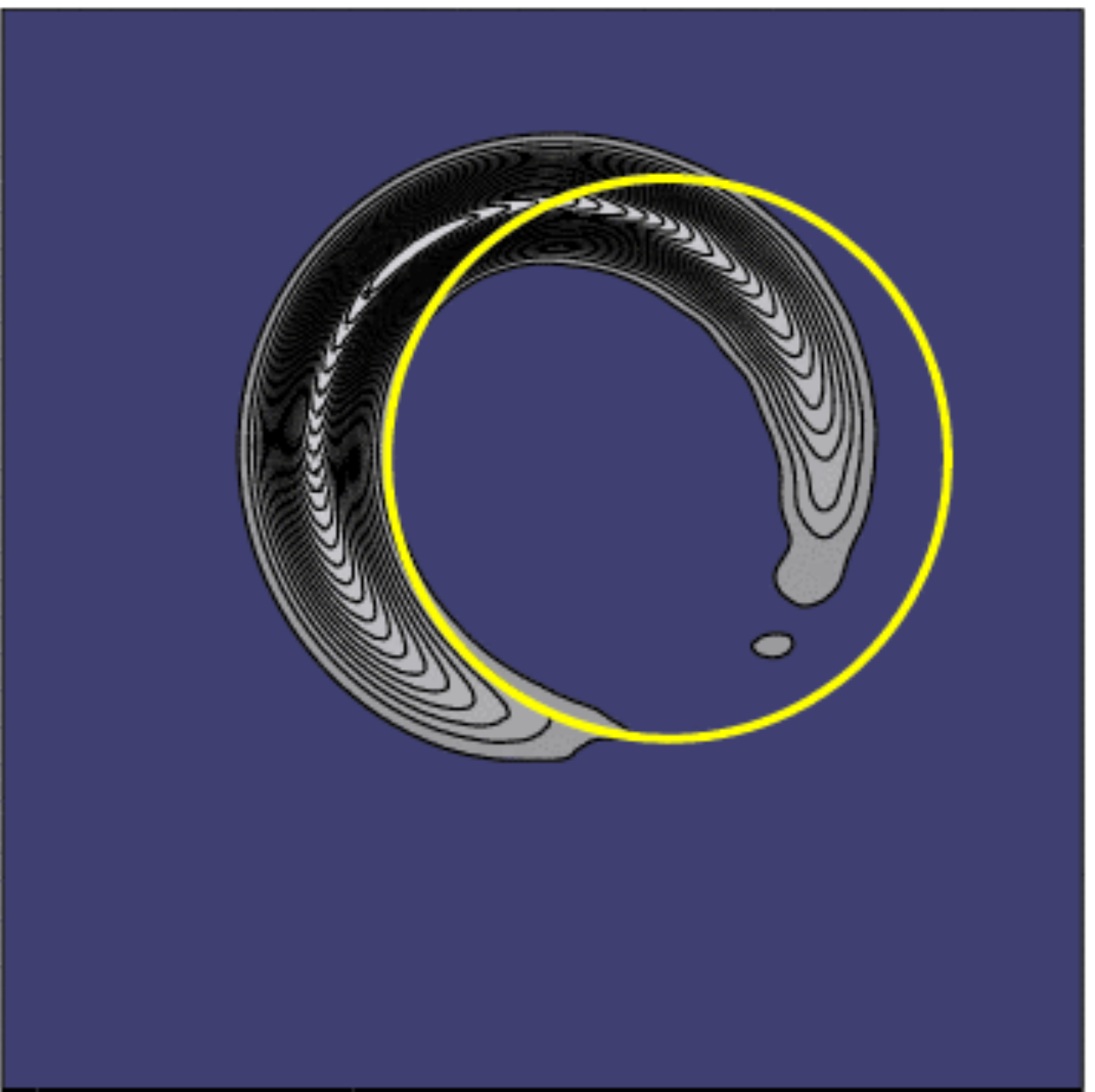}}}
\caption{\label{fig:Spiral_orbit_unif_field} Changes in the shape of the probability density $|\Psi|^2$
during one of the subsequent orbits of the wave packet. The circle represents the classical orbit.
The animation of the dynamics can be accessed on-line \cite{Simi09b}.}
\end{figure}

The complexity of the 
dynamics of the wave packet increased in the later stages of the motion. While the wave packet followed
the circular motion, the probability density $|\Psi|^2$ at some times assumed the spiral shape shown in 
Figure \ref{fig:Spiral_orbit_unif_field}, increased or reduced 
its length, changed its rotational motion, and translated  from one place to another. 
Overall, the characteristics of this dynamics were similar to the characteristics  described in 
Ref. \cite{Schl08} and \cite{Demi08}. While the figures show some of the complex shapes
of the probability density $|\Psi|^2$, the richness of the electron motion can be better appreciated through
the animation of the dynamics accessible on-line \cite{Simi09b}. 

\begin{figure}
\centering
\vspace*{- 10.5 cm}
{\scalebox{.6}{\includegraphics{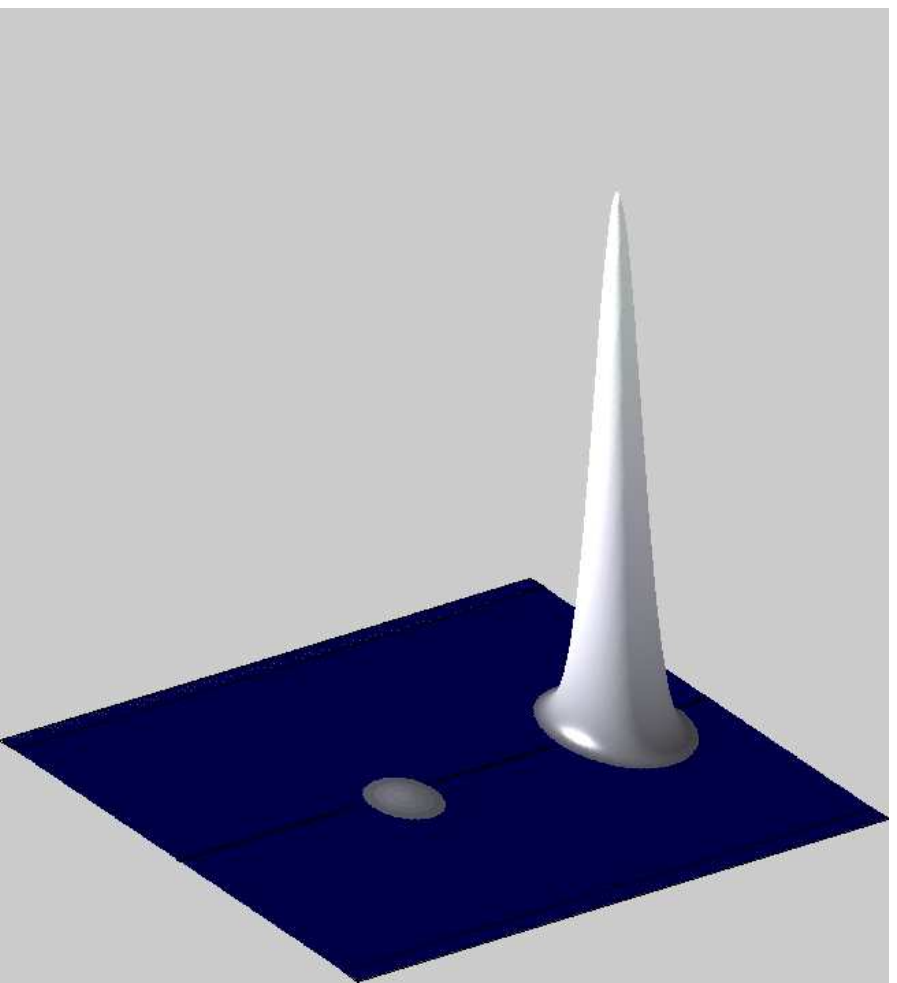}}}\hspace*{- 7.4 cm}
{\scalebox{.6}{\includegraphics{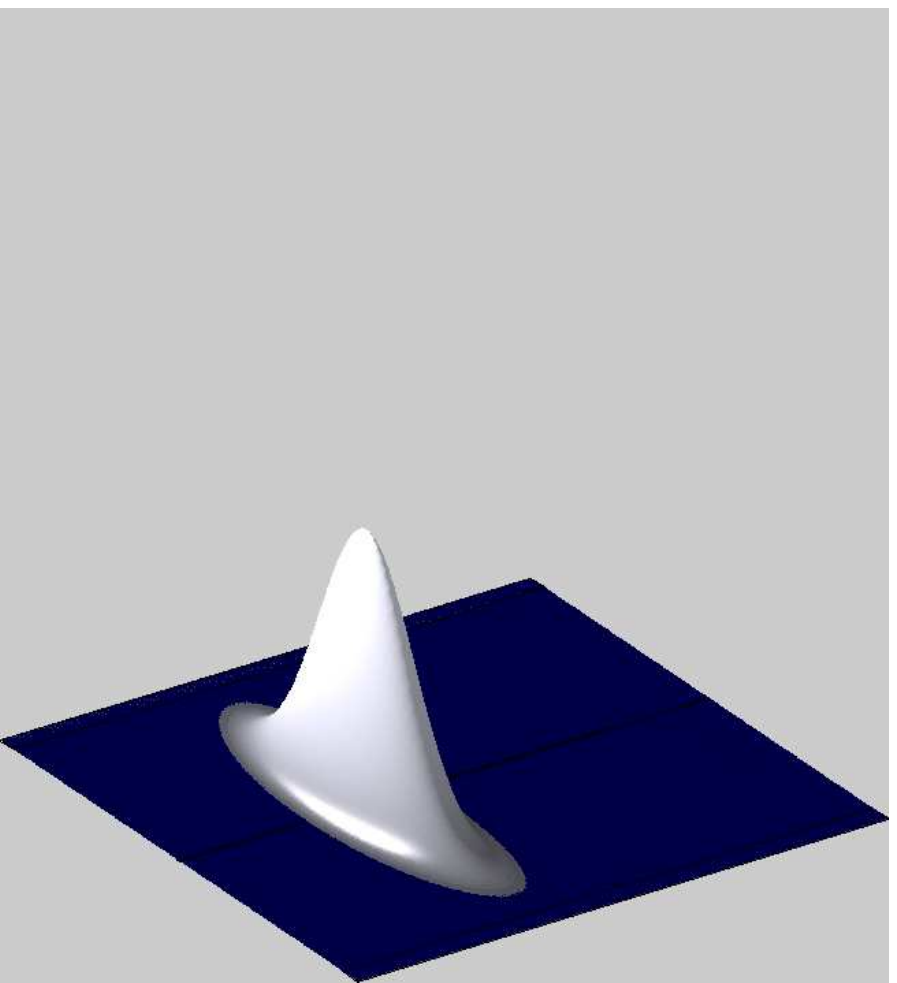}}}\hspace*{- 7.4 cm}
{\scalebox{.6}{\includegraphics{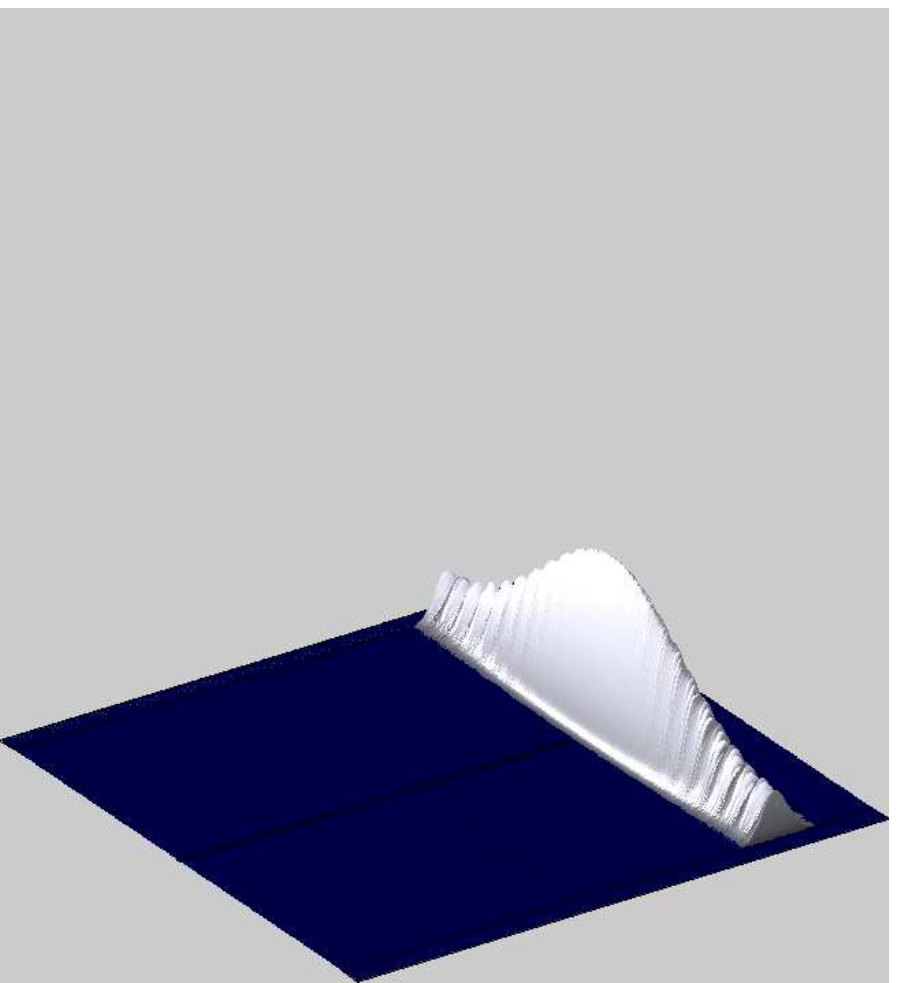}}}\vspace*{- 10.6 cm}
{\scalebox{.6}{\includegraphics{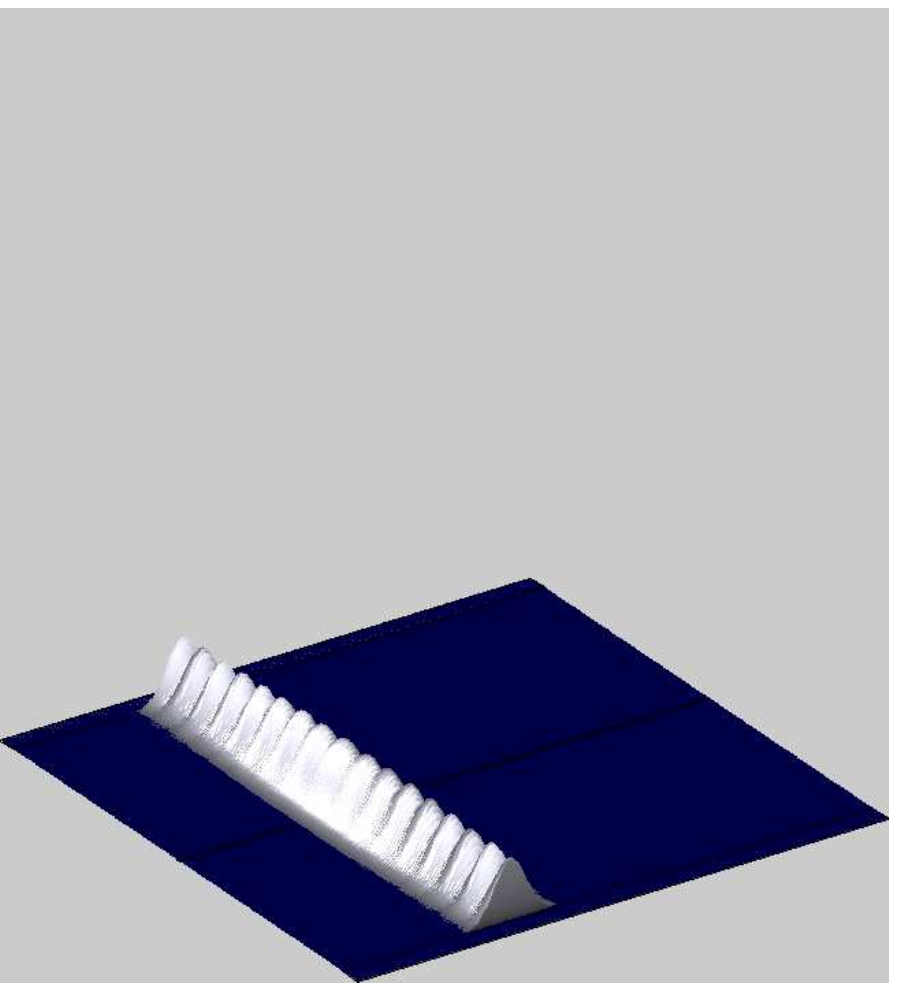}}}\hspace*{- 7.4 cm}
{\scalebox{.6}{\includegraphics{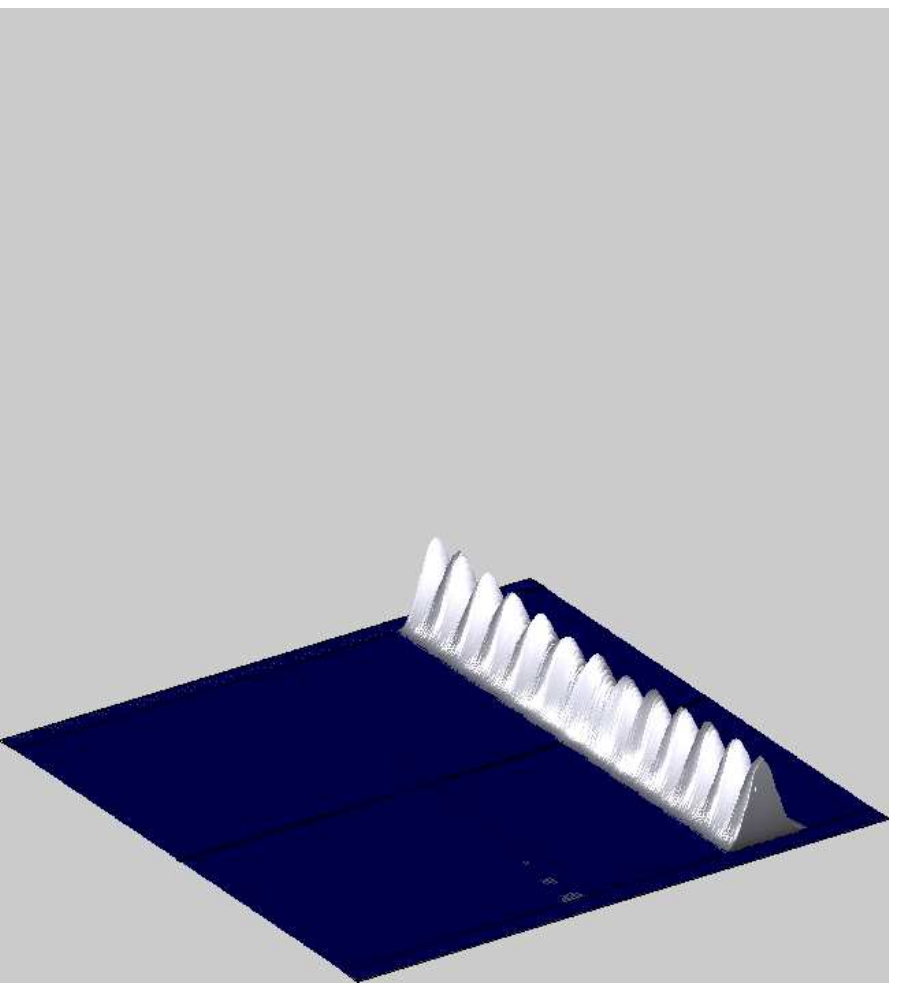}}}\hspace*{- 7.4 cm}
{\scalebox{.6}{\includegraphics{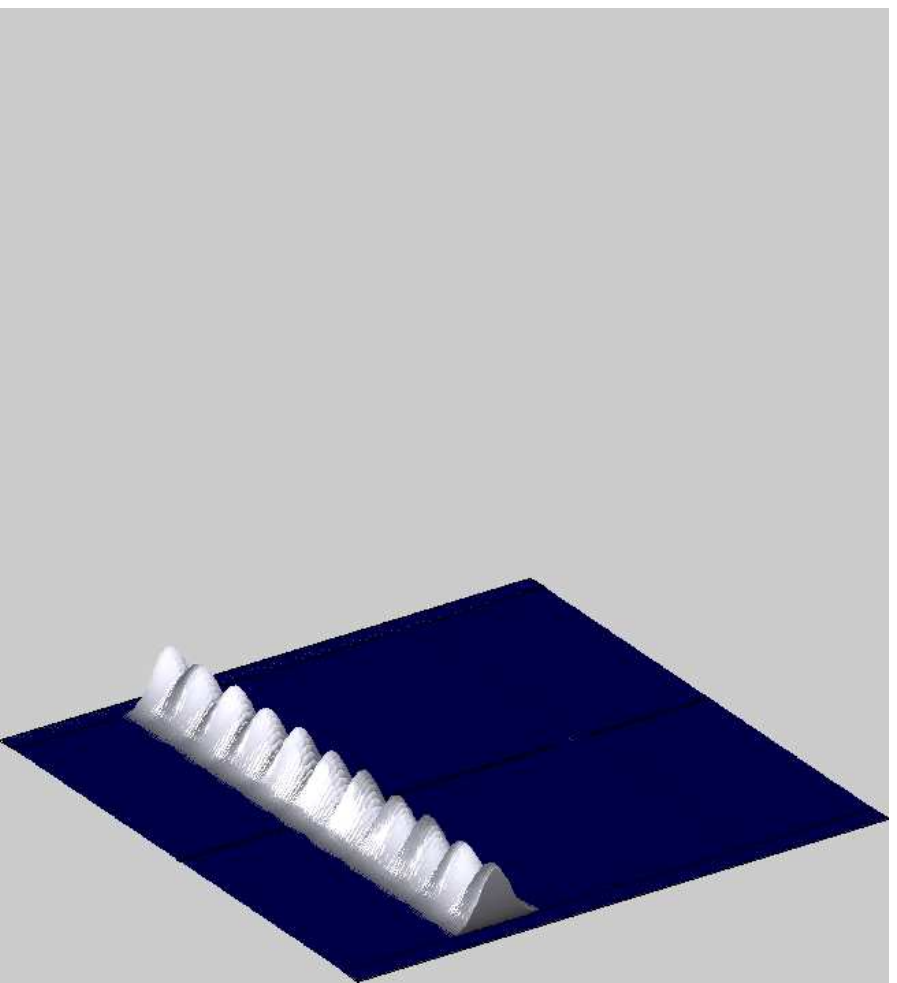}}}\vspace*{- 10.6 cm}
{\scalebox{.6}{\includegraphics{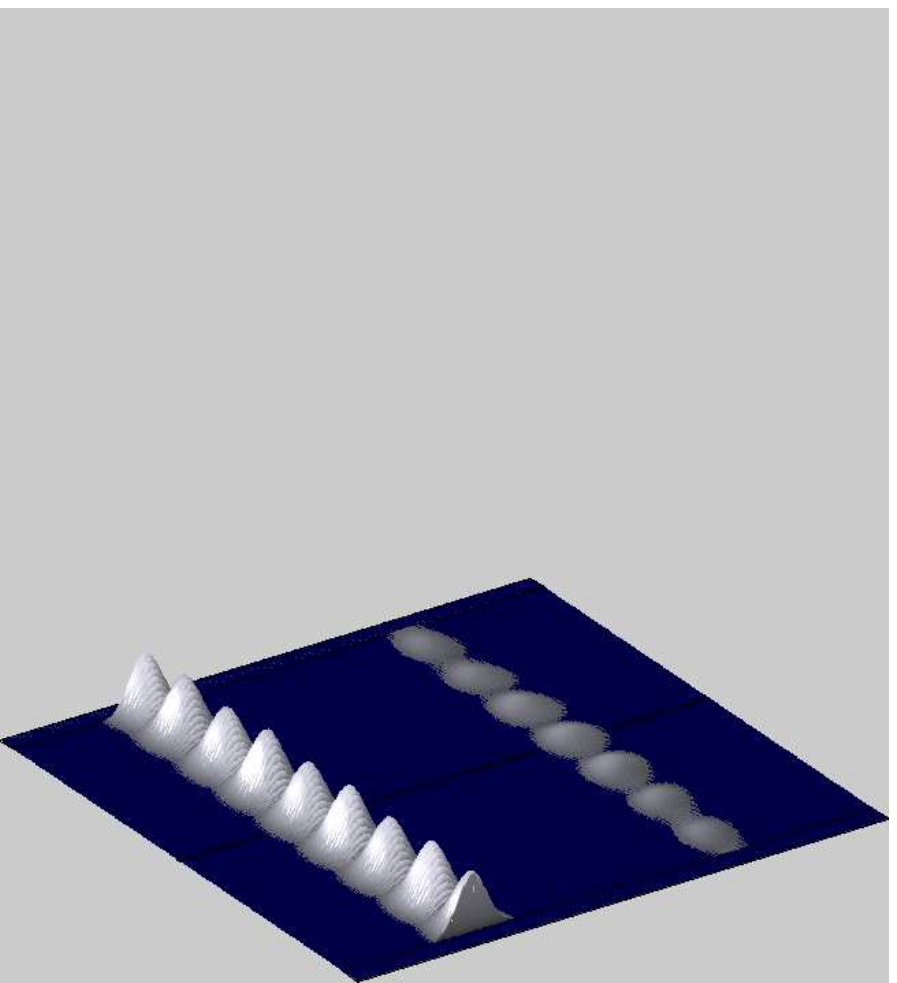}}}\hspace*{- 7.4 cm}
{\scalebox{.6}{\includegraphics{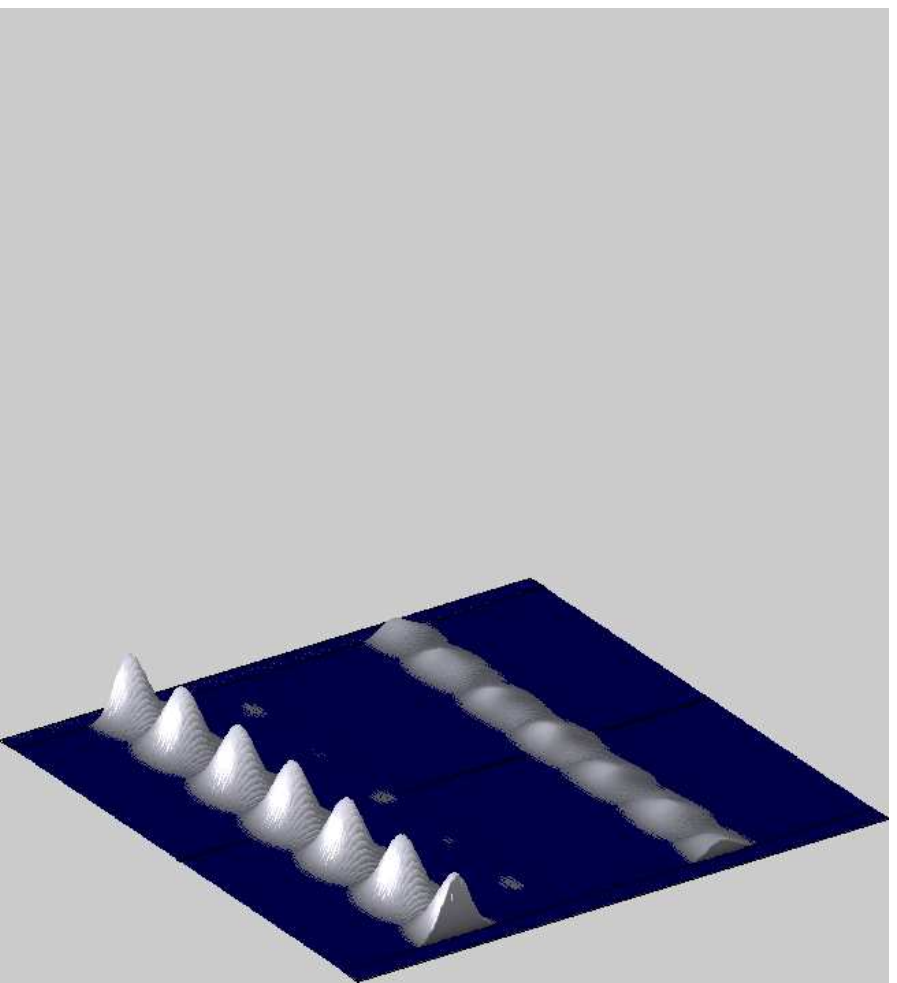}}}\hspace*{- 7.4 cm}
{\scalebox{.6}{\includegraphics{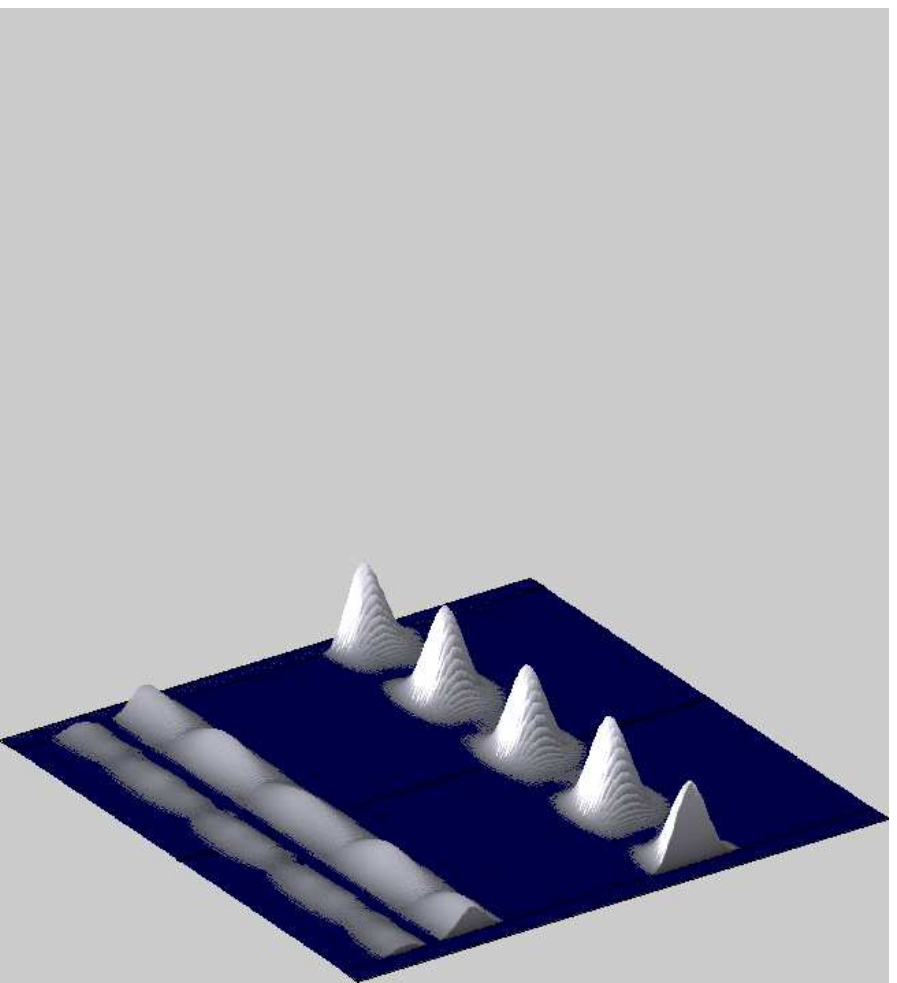}}}
\caption{\label{fig:Circ_mot_vert_cut} The shape of the probability density $|\Psi|^2$ in the vertical plane 
at the time when the wave packet crosses it.  As the wave packet disperses and reaches 
the non-penetrable scalar potential $A_{0}$, the electron wave function creates a bound state 
in the direction of the magnetic field. As the  wave packet rotates in the uniform magnetic field,
the number of nodes reduces.
The related animation can be accessed on-line \cite{Simi09b}.}
\end{figure}

There was no intention
of this validation to study the property of the confinement of the electron by the electrostatic potential. But, 
by using a scalar potential $A_{0}$ at the boundaries of the computational volume to limit the electron motion 
in the direction of the magnetic field, due to the wave packet dispersion, we have inadvertently also 
solved the Dirac equation for the confined electron. We reported \cite{Sim09} that, 
contrary to the claim of the Klein paradox,
there was no penetration of the wave packet into a potential barrier of a 
supercritical potential of height $V$ satisfying the condition $eV>E+mc^{2}$.
The time evaluation of the solution of the Dirac equation shown in Figure \ref{fig:Circ_mot_vert_cut} 
confirms this result. Figure \ref{fig:Circ_mot_vert_cut} shows 
that the electron wave function, as the wave packet disperses and reaches a non-penetrable electrostatic potential, 
creates a bound state. Reduction of number of nodes in a bound state solution could be interpreted as 
a transition of  
the particle from the higher energy state into the lover 
energy state as it orbits in a uniform magnetic field.

Several additional tests of the FDTD method were performed using the dynamics of the 
electron in the uniform magnetic field. As expected, due to the normal spin and the magnetic field orientation,
reversal of the orientation of the 
magnetic field resulted in the reversal of the direction of the rotation of the wave packet, keeping the
same properties of the dynamics of motion. 

Of particular interest was testing the effects of the choice of gauge. 
The motion of the same wave packet was also studied for the 
uniform magnetic field oriented along the y-axis defined by the translationally invariant gauges 
\begin{equation}
\vec A={B_{0}} (-z,0,0),
\label{Uni_mag_potent_x}
\end{equation}
or
\begin{equation}
\vec A={B_{0}} (0,0,x).
\label{Uni_mag_potent_z}
\end{equation}

In both cases the wave packet persisted in a circular motion following the classical orbit. While
the dynamics of the wave packet behaved as expected, using translationally invariant gauges has 
an additional importance on validating the FDTD computation.
As seen in Eq. (\ref{Psi_1}), the gauge in Eq. (\ref{Uni_mag_potent_x}) couples to the $\Psi_{4}$ component
and the gauge in Eq. (\ref{Uni_mag_potent_z}) couples to the $\Psi_{3}$ component. Similarly, the cross coupling 
exists for other components of the wave function $\Psi$. As a result, the same dynamics should be obtained 
by different combinatorics of the components of the vector potential and the components of the wave function. 
This enabled for testing of possible inconsistencies in the FDTD updating equations. No  inconsistencies 
were found.

The dynamics of the wave packet motion in a uniform magnetic field was used here 
only as a validation 
of the FDTD method when a vector potential was applied in the Dirac equation. The same complexity 
of the quantum dynamics of motion as in previous publications \cite{Schl08,Demi08} was shown.
While used here only for computational validation of the FDTD method, this dynamics could
be studied as a separate problem in more detail in the future.
    
Finally, the complexity of the quantum dynamics
of the wave packet in a uniform magnetic field studied here for three choices of gauge
could be fully appreciated only by downloading the related animations \cite {Simi09b}. 
  
\section{Wave packet dynamics in a vector potential created by two infinite solenoids}

The goal of this work is to study the dynamics 
of an electron wave packet under the influence of a vector potential associated 
with the Aharonov-Bohm effect \cite{AB59}. Particularly, in this paper we present 
the dynamics of a wave packet obtained from the solution of the Dirac equation 
with a vector potential created by two infinite solenoids. 

\begin{figure}
\centering
\vspace*{- 0.5 cm}
{\scalebox{.45}{\includegraphics{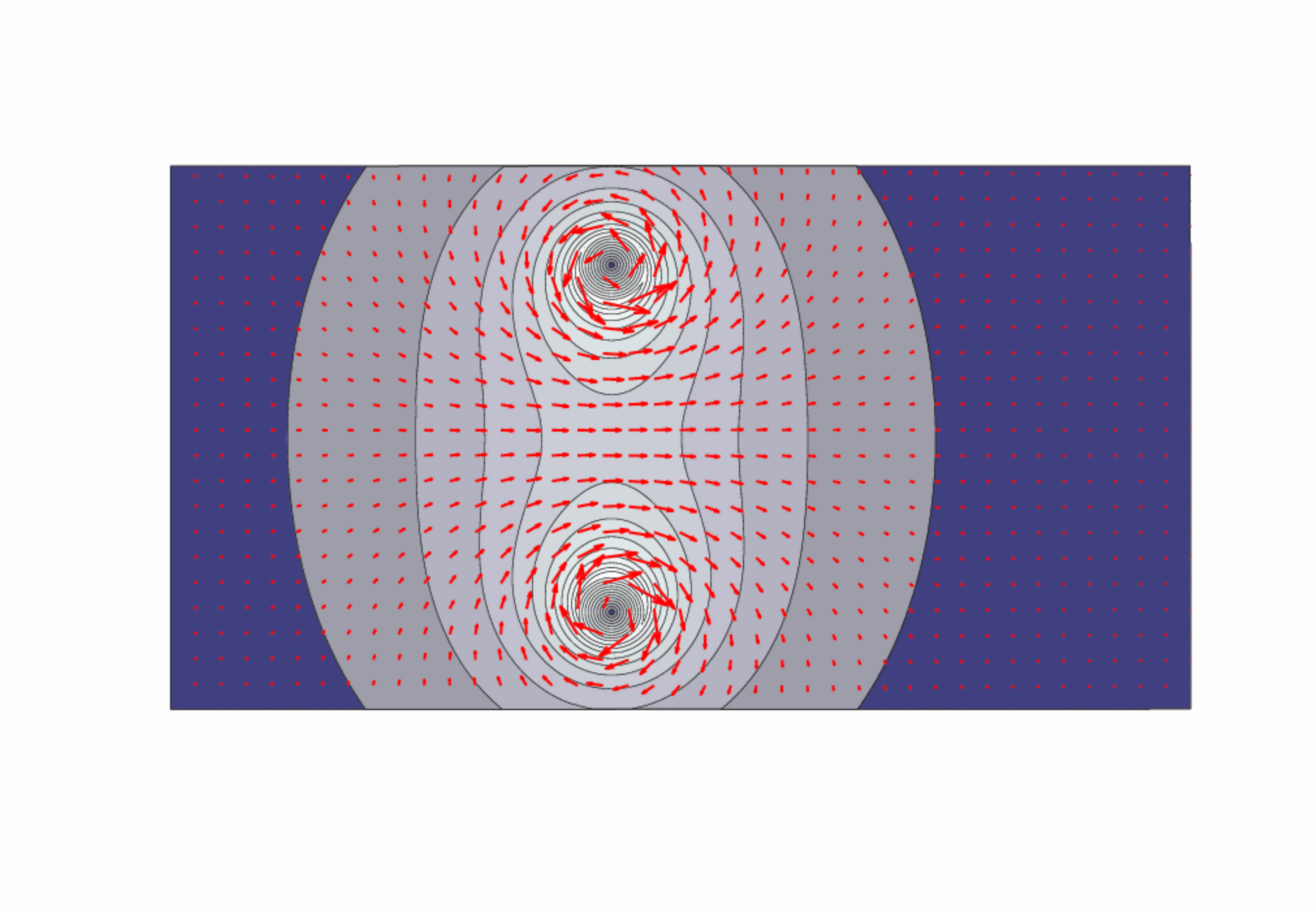}}}
\vspace*{- 2.8 cm}
\caption{\label{fig:plot_of_vector_potential} Vector potential of two parallel infinite solenoids
with oppositely directed magnetic fluxes of the same magnitude. The potential is shown in the plane
normal to the orientation of the solenoids.}
\end{figure}

The vector potential of a single infinite solenoid oriented along the y-axis can be written as

\begin{equation}
{\vec A =\left\{ \begin{array} {c} {\Phi \over {2 \pi R_{0}^{2}}} (-z,0,x) \;\;\;\;\; \mbox{ for $r \leq R_{0}$} \\ 
{\Phi \over {2 \pi r^{2}}}(-z,0,x) \;\;\;\;\; \mbox{ for $r > R_{0}$}
 \end{array} \right. },
\label{Vect_pot_inf_solenoid}
\end{equation}
where $\Phi=B_{0} \pi R_{0}^{2}$ is the magnetic flux inside the solenoid, $B_{0}$ defines the 
strength of the magnetic field, $R_{0}$ is the radius of the solenoid,  and
$r=\sqrt{x^{2} + z^{2}}$ is the distance from the center of the solenoid in the x-z plane. Outside two 
parallel infinite solenoids separated by a distance $2a$ and with opposite magnetic field 
orientation, the vector potential is then

\begin{equation}
\! \! \! \! \! \! \! \! \! \! \! \! \! \! \! \! \! \! \vec A= {\Phi \over {2 \pi}} \left ( {{z+a} \over {x^{2} + (z+a)^{2}}}- {{z-a} \over{x^{2} + (z-a)^{2}}}
,0,{{x} \over{x^{2} + (z-a)^{2}}}-{{x} \over{x^{2} + (z+a)^{2}}}\right ) \!  .
\label{Vect_pot_two_inf_solenoid}
\end{equation}
An example of the shape of this potential is shown in Figure \ref{fig:plot_of_vector_potential}.
Outside the solenoids, the associated magnetic fields $\vec B = \vec \nabla \times \vec A$ 
of the vector potentials described by Eqs. (\ref{Vect_pot_inf_solenoid}) 
and (\ref{Vect_pot_two_inf_solenoid}) are zero. 

The time-dependent Dirac equation is now solved for the scalar component 
of the four-potential $A_{0}=0$ and the vector component $\vec A$ defined by  
Eq. (\ref{Vect_pot_two_inf_solenoid}). With this choice  we  can study the dynamics of the 
wave packet in the region where the electric and the magnetic fields are zero. Obviously, 
in this region the Lorentz force acting on the electron, $\vec F = e (\vec E + \vec v \times \vec B)$, is also zero.
 
The dynamics of the electron motion between two infinite and parallel solenoids is essentially the same as the
electron dynamics in the case of Aharonov-Bohm effect. In both cases the electron moves in a field free region 
with no Lorentz force acting on it. Because of the symmetry, however, the dynamics in the case of 
two infinite solenoids consists 
only of the motion along one straight line between the solenoids. This avoids the complications of 
the  Aharonov-Bohm dynamics where the trajectories on opposite sides of a single solenoid are compared.  
Because the properties of the dynamics are not studied through an interference, 
in the case of two solenoids the changes
in the dynamics of the electron wave packet exclude the contribution of the nonlocal properties of the electron 
wave function dependent on the topology of the space. Another advantage of using two solenoids 
and localized wave packet is the possibility to separate the solenoids far enough that no parts of 
the wave packet penetrate the non-zero field region inside the solenoids, excluding this contribution to the dynamics 
as well. If necessary, the solenoids could be shielded by a potential barrier of a supercritical potential 
additionally preventing any penetration of the wave packet into non-zero field region.

\begin{figure}
\centering
\vspace*{- 9. cm}
{\scalebox{.55}{\includegraphics{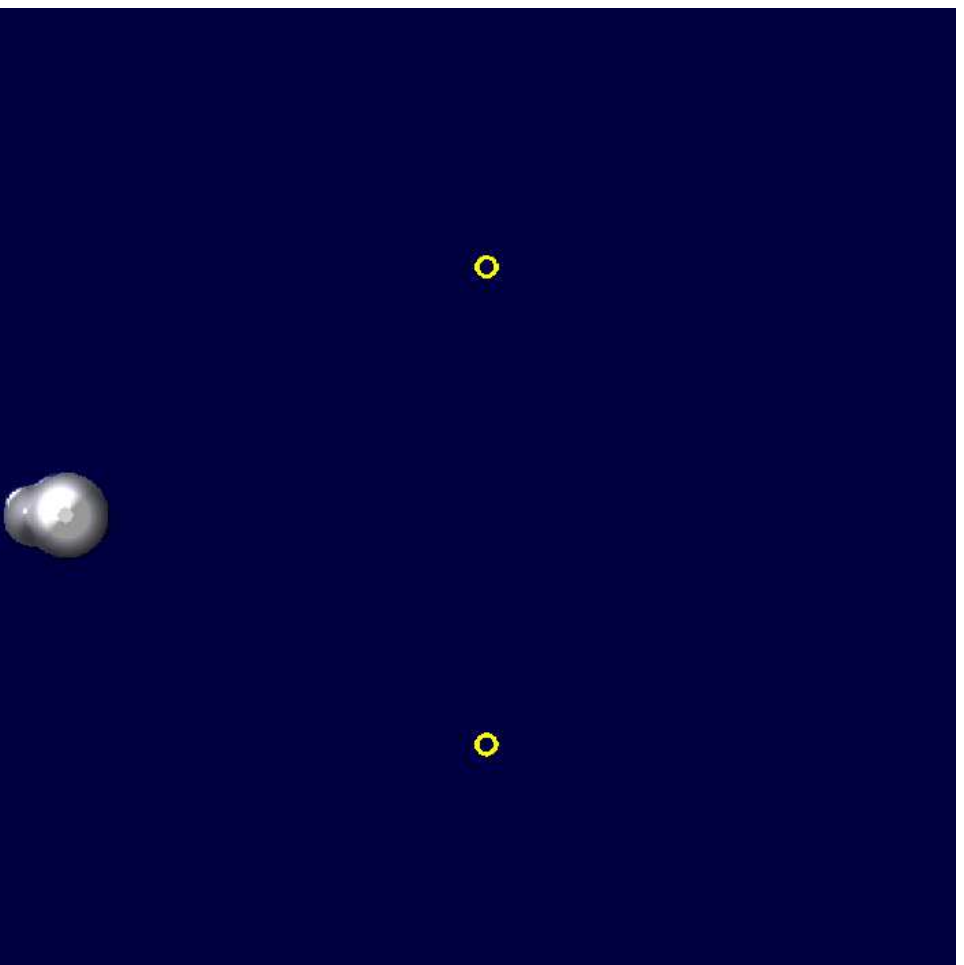}}}\hspace*{- 6.5 cm}
{\scalebox{.55}{\includegraphics{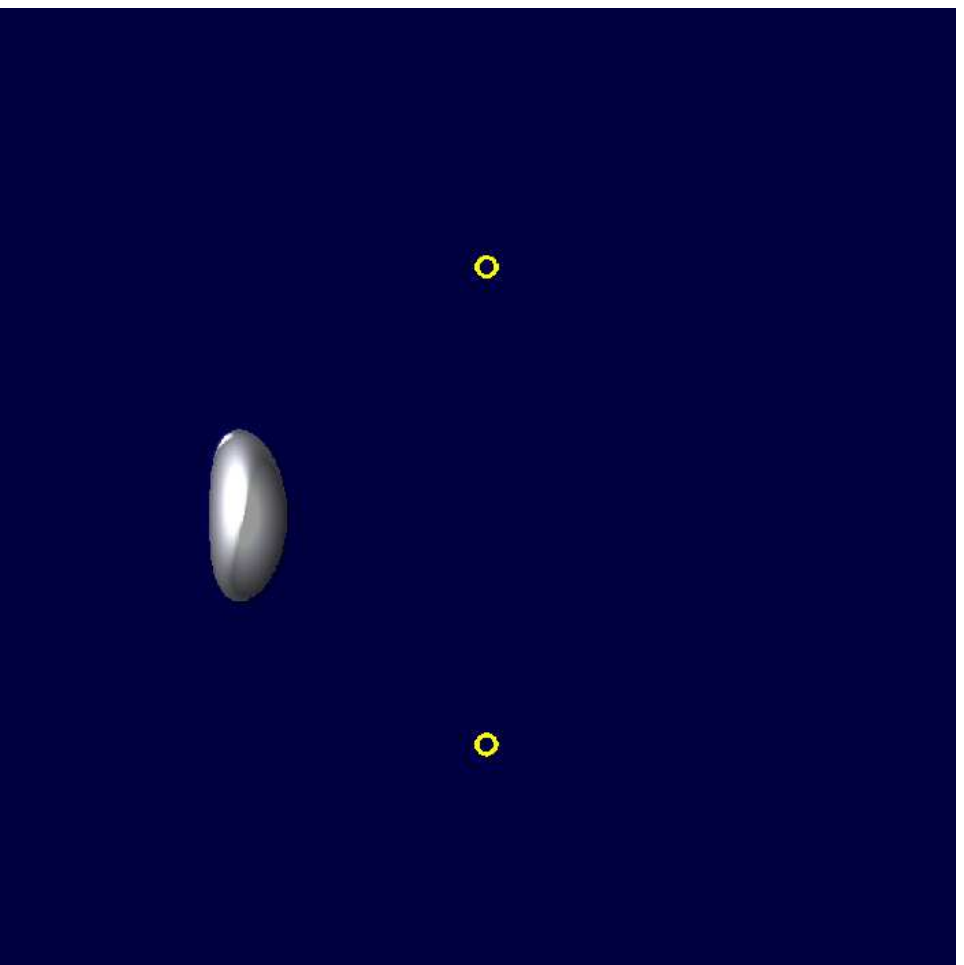}}}\hspace*{- 6.5 cm}
{\scalebox{.55}{\includegraphics{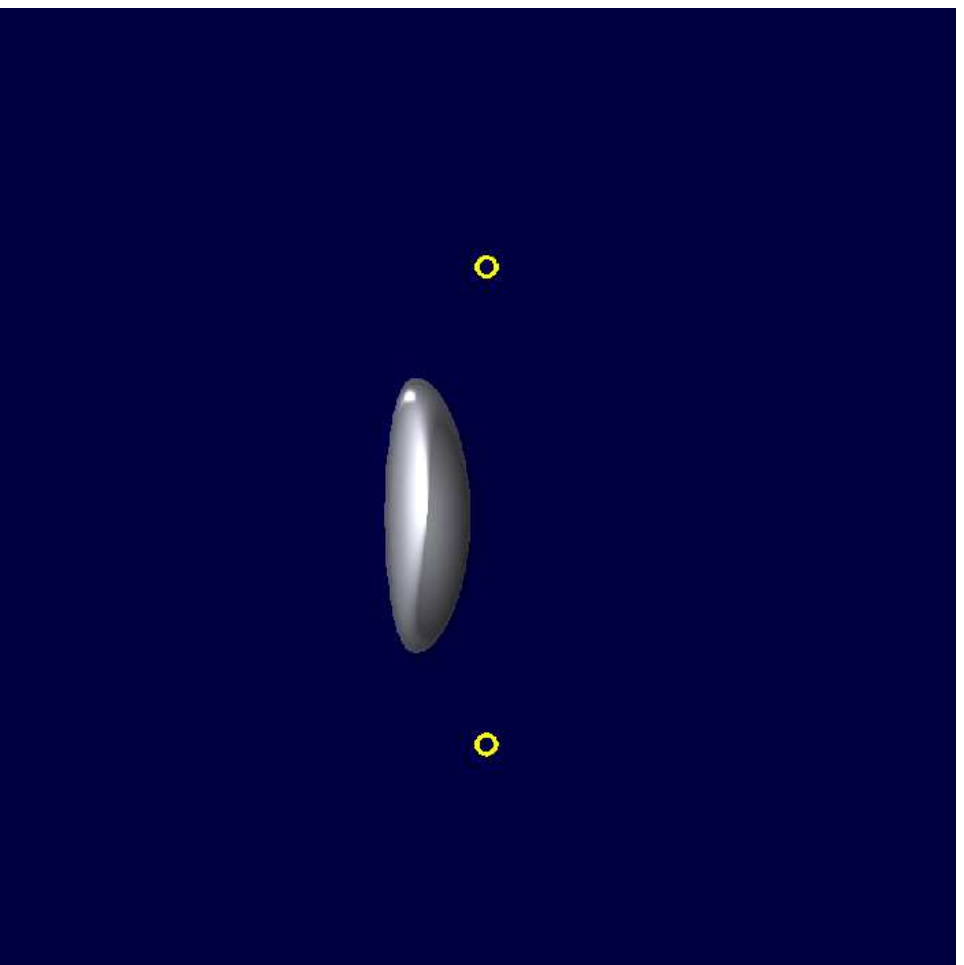}}}\vspace*{- 9.75 cm}                   
{\scalebox{.55}{\includegraphics{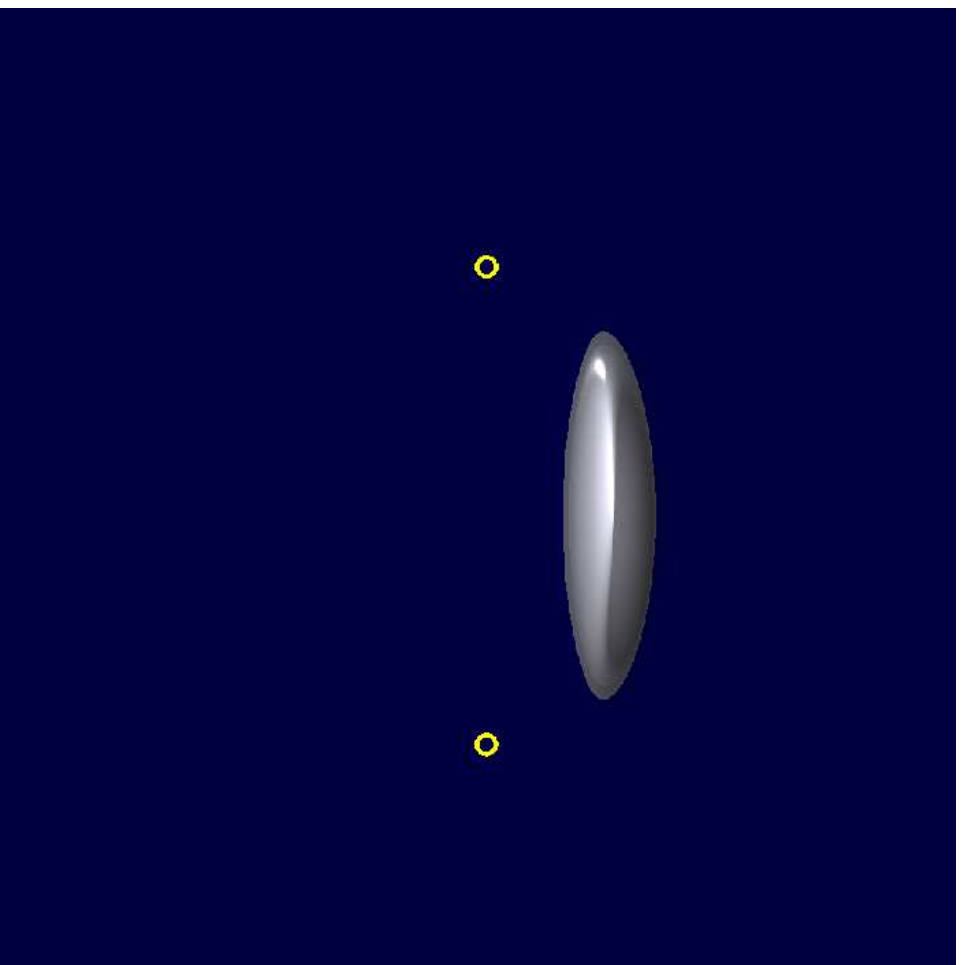}}}\hspace*{- 6.5 cm}
{\scalebox{.55}{\includegraphics{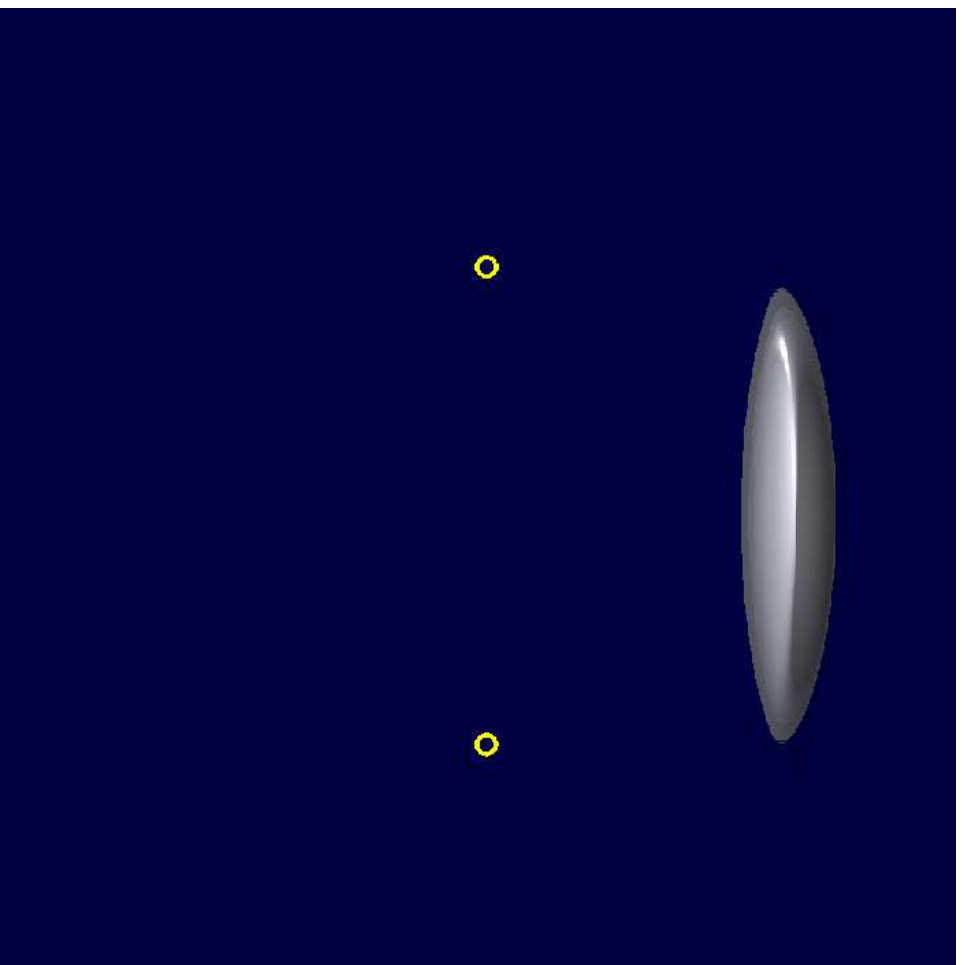}}}\hspace*{- 6.5 cm}
{\scalebox{.55}{\includegraphics{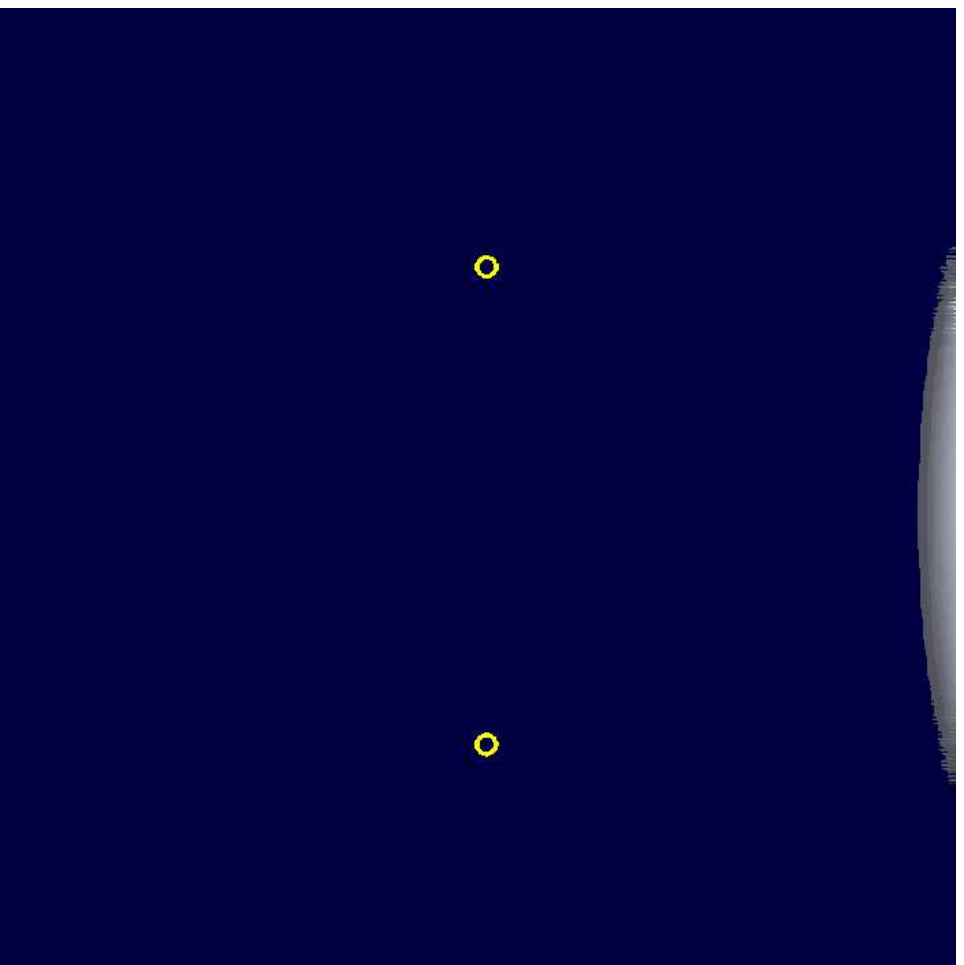}}}
\caption{\label{fig:Wave_two_solenoids} Six stages of the motion of
the wave packet between two infinite parallel solenoids.
Small yellow circles represent the positions and the sizes of the solenoids.
The animation can be accessed on-line \cite{Simi09b}.}
\end{figure}

The motion of the wave packet between the solenoids, in the plane
normal to the orientation of the solenoids, is shown in Figure \ref{fig:Wave_two_solenoids}. 
The wave packet was initially positioned away from center of the solenoids. The dynamics was 
studied for two initial momenta, $p_{1} = 0.53 \; MeV/c$  or $p_{1} = 0.64 \; MeV/c$. 
The solenoids were separated by a distance of $2a=0.1 \; nm$ and the magnitude of the magnetic flux 
inside  each of  them was $\Phi=5.2 \times 10^{-14} \; Wb$.

\begin{figure}
\centering
\vspace*{- 6. cm}
{\scalebox{.75}{\includegraphics{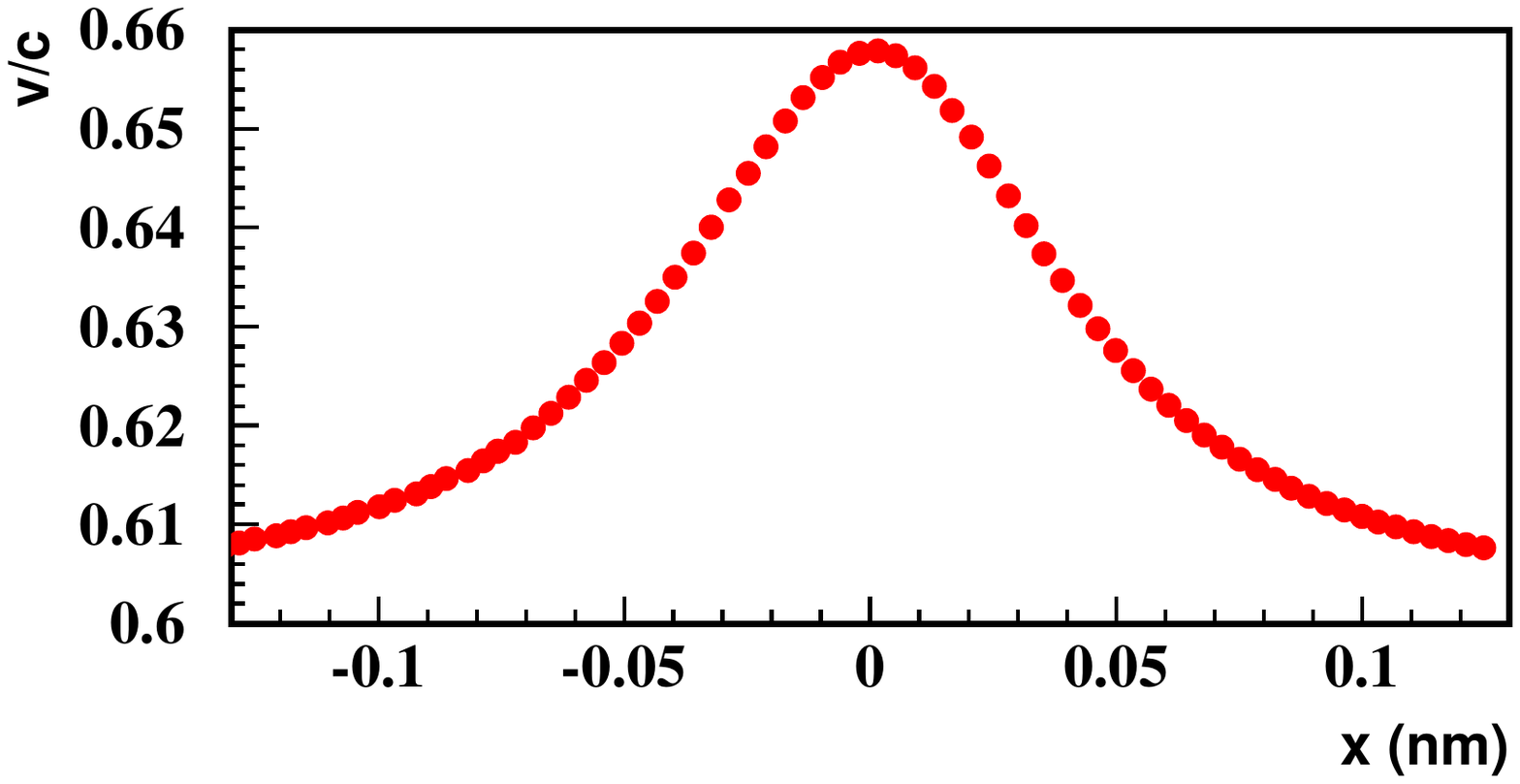}}}
\vspace*{- 7.5 cm}
\caption{\label{fig:two_solenoids_velocity} The velocity, as a function of position, of a wave packet 
moving along a straight line between two infinite parallel solenoids normalized to the speed of light . 
The solenoids are positioned at x=0.} 
\end{figure}

As shown in Figure \ref{fig:Wave_two_solenoids}, and in the related animation, the wave packet 
moved between two solenoids along a straight line, dispersing in time in the direction 
normal to the direction of motion. 
Since, classically, there was no force acting on the electron, one should have expected a constant velocity 
of the wave packet along the straight line. This was not the case. The velocity
of the wave packet, shown in Figure \ref{fig:two_solenoids_velocity} as a function of the position,
increased as the packet approached the solenoids and decreased as the packet 
left the solenoids. It is somewhat paradoxical that while there was, classically, no force acting on the electron, 
the electron acceleration, obtained strictly as a solution of the Dirac equation, was not zero. 
In the next section we will show that this paradox can be resolved.
 
\section{The dynamics of the charged particle under the conditions of the Aharonov-Bohm effect}

The controversy in the previous section, resulting from the solution of the Dirac equation,
can be understood if we look again at the basic principles of classical mechanics.

The motion of a particle in the absence of a force in classical dynamics is obtained by solving 
the differential equation
\begin{equation}
{d\vec p \over dt}=0,
\label{Newton_no_force}
\end{equation}
where $\vec p=m\vec v$ is the momentum of a particle. The solution of this equation results 
in a constant velocity of the particle. The solution of the Dirac equation
shows that, even if the force acting on the particle is zero, the velocity of the particle changes. To 
resolve the discrepancy between classical dynamics and the Dirac equation we need to look at the basics of
how the Dirac equation of an electron is obtained.

In obtaining the Dirac equation from classical dynamics, the classical momentum $\vec p$ 
becomes a momentum operator $\hat{p}= -i\hbar \vec \nabla$. If the electron is in a given external field 
the momentum operator is replaced by $\hat{p} \rightarrow -i\hbar \vec \nabla + |e| \vec A$, where $e$ is 
the charge of the electron \cite{Bere82}. This replacement not only is not self-evident, but must be 
done in the first order equations \cite{Bere82}. Going backward from the Dirac equation into classical dynamics, 
one replaces the classical momentum $\vec p \rightarrow \vec p + |e| \vec A$. In the Lagrangian mechanics, 
the momentum $\vec p = m\vec v + |e| \vec A$ is called the generalized or canonical momentum \cite{Semon96}. 
By proper use of momentum conservation \cite{Semon96}, the motion of an electron in the absence 
of a classical force can now be  obtained by solving the differential equation
\begin{equation}
{d{(\vec p + |e| \vec A)} \over dt}=0.
\label{general_no_force}
\end{equation}
Since $A_{x}$ does not explicitly depend on time, for the motion along the x-axis,
Eq. (\ref{general_no_force}) can be written as
\begin{equation}
{d p_{x} \over dt} =  - |e| {d A_{x} \over dt}=- |e| {d A_{x} \over dx}{dx \over dt}=- |e| {d A_{x} \over dx}v_{x}
\label{component_general_no_force}
\end{equation}
In the relativistic case 
\begin{equation}
{d p_{x} \over dt} =\left( 1-{v_{x}^{2} \over c^{2}}\right)^{-{3 \over 2}} m_{0}  {d v_{x} \over dt}. 
\label{rel_momentum}
\end{equation}
Here $m_{0}$ is the rest mass of the electron.  Using Eqs. (\ref{general_no_force}) and (\ref{rel_momentum})
we get the differential equation for the velocity of the electron
\begin{equation}
{d v_{x} \over dt} =- {|e| \over m_{0}} v_{x} \left( 1-{v_{x}^{2} \over c^{2}}\right)^{3 \over 2}
{d A_{x} \over dx}. 
\label{rel_diff_eq}
\end{equation}
Due to the relativistic motion, in the system of the solenoids the velocity differential 
$d v_{x}$ has to be substituted with
 $d v_{x}/(1-v_{x}^{2}/c^{2})$. This  leads to the final form of the differential equation
\begin{equation}
{d v_{x} \over dt} =- {|e| \over m_{0}} v_{x} \left( 1-{v_{x}^{2} \over c^{2}}\right)^{5 \over 2}
{d A_{x} \over dx}. 
\label{rel_diff_eq2}
\end{equation}
Substituting $A_{x}$ from the Eq.  (\ref{Vect_pot_two_inf_solenoid}) for $z=0$, electron motion along 
the mid-path between two solenoids, the differential equation for the 
velocity of the electron becomes
\begin{equation}
{d v_{x} \over dt} ={{2 |e| \Phi} \over {\pi m_{0}}} { x  \over {(x^{2}+a^{2})^{2}}}
v_{x} \left( 1-{v_{x}^{2} \over c^{2}}\right)^{5 \over 2}, 
\label{rel_diff_eq_final}
\end{equation}
where $a$ is half the distance between two parallel infinite solenoids. This differential equation can be
solved numerically.

\begin{figure}
\centering
{\scalebox{.75}{\includegraphics{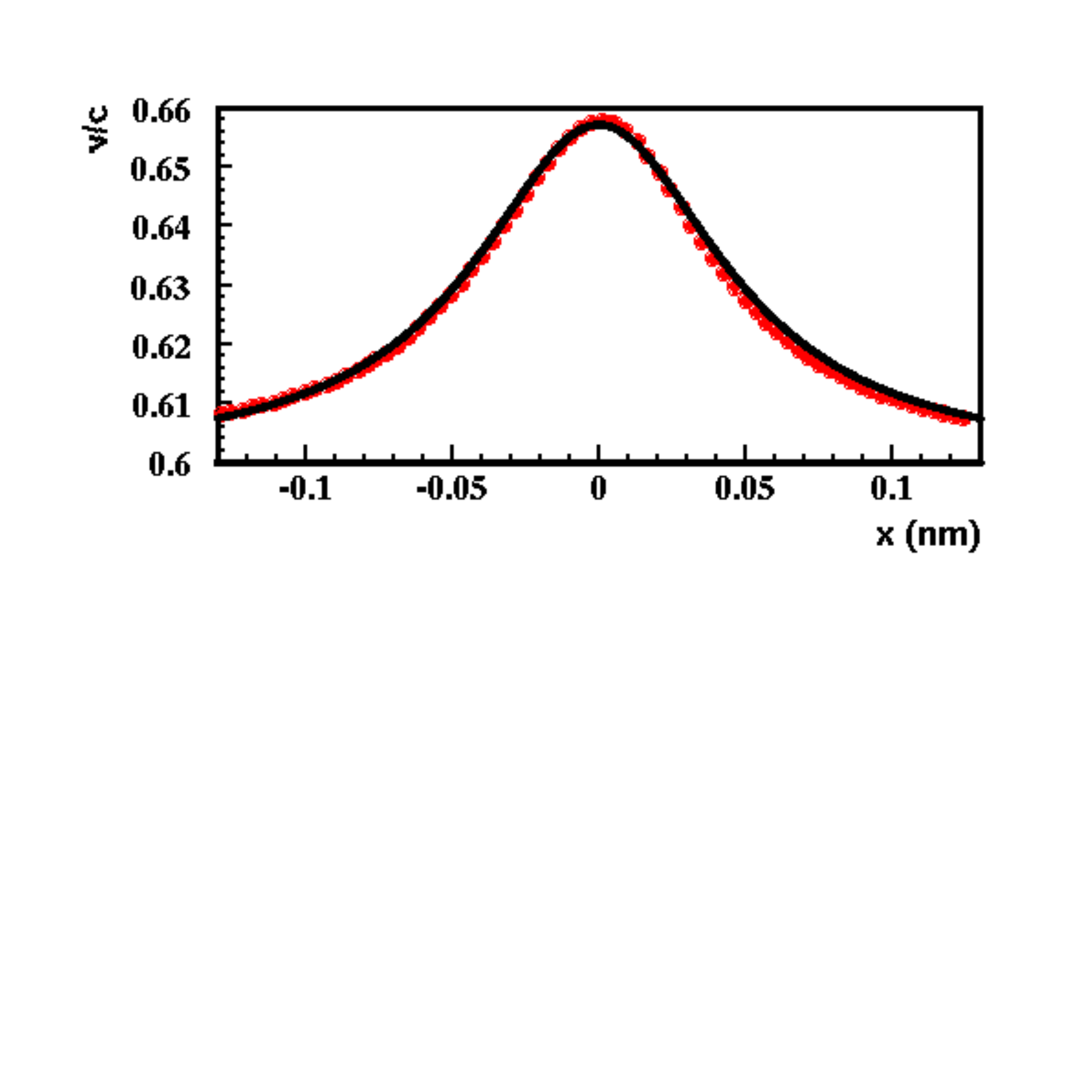}}}
\vspace*{- 8. cm}
\caption{\label{fig:theory_two_solenoids_velocity} The velocity, as a function of position, of a wave packet 
moving along a straight line between two solenoids normalized to the speed of light. The circles represent
the solution of the Dirac equation and the line represents the solution of  Eq. (\ref{rel_diff_eq_final}) under the
same conditions. The solenoids are positioned at x=0.} 
\end{figure}

Figure \ref{fig:theory_two_solenoids_velocity} shows a nearly perfect match between the solution of 
Eq. (\ref{rel_diff_eq_final}) and the solution of the Dirac equation. A small difference, a fraction of
a percent, is the result 
of the numerical precision and the fact that in classical dynamics we calcute propagation of 
a particle, not a wave packet, therefore having no contribution of the localization to the initial condition. 

In a little digression from the main point of this paper, we can show that the above application of 
the conservation of generalized or canonical momentum is not an exception. One can easily show that 
for the translationally invariant gauges, Eqs.  (\ref{Uni_mag_potent_x}) and  (\ref{Uni_mag_potent_z}),
the conservation of generalized or canonical momentum is equivalent to the Lorentz force. Without 
changing the general result, for ease of calculation, we assume that the electric field $\vec E =0$ 
and the vector potential $\vec A$ is time independent. 

The conservation of generalized or canonical momentum, Eq. (\ref{general_no_force}),
for the dynamics of an electron in a uniform magnetic field defined by the translationally invariant gauges,
can now be rewritten as
\begin{eqnarray}
{d{\vec p} \over dt}&=&-|e|{d{\vec A} \over dt}=-|e| (\vec v \vec \nabla) \vec A
=-|e| (\vec v \vec \nabla) (-B_{0} z \; \vec i + B_{0} x \; \vec k) \nonumber \\
&=& -|e| (-B_{0} v_{z} \; \vec i + B_{0} v_{x} \; \vec k)=-|e| (\vec v \times \vec B)
\label{general_Lorentz}
\end{eqnarray}
Eq.  (\ref{general_Lorentz}) shows that the Lorentz force is just a consequence of  
generalized or canonical momentum conservation. While not in such an explicit form, similar results were
obtained by Semon and Taylor \cite{Semon96}.

In a somewhat different way the same can be shown for the rotationally invariant gauge, 
Eq.  (\ref{Uni_mag_potent}).  In this case we use conservation of the canonical angular momentum 
\begin{equation}
{d{(\vec r \times \vec p + |e| \; \vec r \times \vec A)} \over dt}={d{(\vec l + |e| \; \vec r \times \vec A)} \over dt}=0.
\label{general_angular}
\end{equation}
$\vec l=\vec r \times \vec p$ represents classical angular momentum. Eq. (\ref{general_angular}) can be 
rewritten as
\begin{eqnarray}
{d\vec l \over dt}&=&-|e| {d{(\vec r \times \vec A)} \over dt}
=-|e| \left({d\vec r \over dt} \times \vec A + \vec r \times{d\vec A \over dt}\right) \nonumber \\
&=&-|e| \left( \vec v \times \vec A + \vec r \times (\vec v \vec \nabla) \vec A \right)
\label{general_angular2}
\end{eqnarray}
Repeating the same algebra, but using the rotationally invariant gauge of 
the vector potential $\vec A$ , we get
\begin{equation}
{d\vec l \over dt} =-|e| \; \vec r \times ( \vec v \times \vec B) = \vec r \times \vec F.
\label{general_angular3}
\end{equation}
$\vec F$ is again the Lorentz force. With this we have shown the equivalence of the conservation of 
generalized or canonical momentum and canonical angular momentum, and the action of Lorentz force for all
vector potential gauges of the uniform magnetic field.  

To conclude this section, we can look at the ongoing question if the  Aharonov-Bohm effect is the result of 
a force changing the  velocity of a particle passing on opposite 
sides of a single infinitely long solenoid, or there is only a quantum-mechanical phase shift 
\cite{Zhu90, Shel98, Keat01,Gronn06,Hors05,Hors07}. Quantum-mechanically
such a question does not exist. The dynamics of the relativistic electron should be obtained only as 
a solution of the time-dependent Dirac equation. The solution of the Dirac equation shows that the 
velocity of a wave packet changes even in the region where the magnetic field is zero. 
Since the change of the velocity depends on the gradient of  the vector potential, the velocity of the 
wave packet passing on the opposite sides of the solenoid will be different.
Giving to the solution of the Dirac equation a classical picture, the requirement of 
the conservation of the canonical momentum in the region where the magnetic field is zero but
the vector potential is not zero, also results in a velocity change of a charged particle. No mechanical forces 
need to be present in order for the velocity of the particle to change. 
From both, the solution of the Dirac equation and 
from the equivalence of the conservation of  canonical momentum and the action of the Lorentz-force, one can
also see that no work was done on the particle. The particle exits with the same energy regardless on
which side of the solenoid it passed. 

From the studies presented in this paper, the classical picture of the Aharonov-Bohm effect
consists of the phase shift in the wave function of the particles passing on opposite sides of a solenoid
being attributed to time lag resulting from different evolution of the velocities of the particles,
without any involvement of mechanical forces and any change in energy. 
A phase shift then results in an interference pattern.

\section{Conclusion}

In conclusion, the full three-dimensional Finite Difference Time Domain (FDTD)
method was developed to solve the Dirac equation. In this paper, the method was
applied to the dynamics of the electron wave packet in a vector potential
in the arrangements associated with the Aharonov-Bohm effect.
The solution of the Dirac equation showed that the velocity of the electron wave packet 
changed even in the region where the electric and the magnetic fields were zero, and 
therefore no force acted on the electron. 

The solution of the Dirac equation agreed
with the prediction of classical dynamics under the assumption that the dynamics were defined by 
the conservation of generalized or canonical momentum. It was shown that in the case of a uniform
magnetic field, the conservation of generalized or canonical momentum was 
equivalent to the action of the Lorentz force.

The studies in this paper have helped to establish a classical picture of the Aharonov-Bohm effect as 
the interference pattern resulting from the phase shift in the wave function of the particles passing 
on opposite sides of a solenoid attributed to a time lag resulting from different evolution 
of the velocities of the particles. No mechanical forces need to be involved and no change in energy
of the particle occurs.

\section*{Acknowledgments}

I would like to thank Dentcho Genov, B. Ramu Ramachandran, Lee Sawyer, Ray Sterling and Steve Wells 
for useful comments.
Also, the use of the high-performance computing resources provided by Louisiana Optical
Network Initiative (LONI; www.loni.org) is gratefully acknowledged.


\section*{References}

\begin{thebibliography}{18}

\bibitem{Sim08} Simicevic N 2008 arXiv:0812.1807v1 [physics.comp-ph]
\bibitem{Sim09} Simicevic N 2009 arXiv:0901.3765v1 [quant-ph]
\bibitem{Yee66}Yee K S 1966 {\it IEEE Trans. Antennas Propagat.} {\bf AP-14} 302
\bibitem{Klein29} Klein O 1929  {\it Z. Phys.}  {\bf 53} 157
\bibitem{AB59} Aharonov Y and Bohm D 1959 {\it Phys. Rev.}  {\bf 115} 485
\bibitem{Cham60} Chambers R G 1960 {\it Phys. Rev. Lett.} {\bf 5} 3
\bibitem{Tono86} Tonomura A, Osakabe N, Matsuda T, Kawasaki T, Endo J, Yano S, and Yamada H 1986
{\it Phys. Rev. Lett.} {\bf 56} 792
\bibitem{Osak86} Osakabe N, Matsuda T, Kawasaki T, Endo J, Tonomura A, Yano S, and Yamada H
{\it Phys. Rev. A} {\bf 34} 815
\bibitem{Pesh89} Peshkin M and Tonomura A 1989 {\it The Aharonov-Bohm effect (Lecture Notes in Physics} vol 340)
(Berlin: Springer). 
\bibitem{Hege08} Hegerfeldt G C and Neuman J T 2008 {\it J. Phys. A: Math. Theor.} {\bf 41} 155305
\bibitem{Boy06} Boyer T H 2006 {\it J. Phys. A: Math. Gen.} {\bf 39} 3455
\bibitem{Boy08} Boyer T H 2008 {\it Found. Phys.} {\bf 38} 498
\bibitem{Cap07} Caprez A, Barwick B, and Batelaan H 2007  {\it Phys. Rev. Lett.} {\bf 99} 210401
\bibitem{Grein85} Greiner W, Muller B, and Rafelski J 1985 {\it  Quantum Electrodynamics 
of Strong Fields}(Berlin: Springer-Verlag).
\bibitem{Sak87} Sakurai J J 1987 {\it  Advanced Quantum Mechanics}
(Redwood City: Addison-Wesley).
\bibitem{Huang52} Huang K 1952 {\it Am. J. Phys.}  {\bf 20} 479
\bibitem{Schl08} Schliemann J 2008  {\it Phys. Rev. B} {\bf 77} 125303
\bibitem{Demi08} Demikhovskii V Ya, Maksimova G M, and Frolova E V 2008  {\it Phys. Rev. B} {\bf 78} 115401
\bibitem{Simi09b} Simicevic N 2008 {\it http://caps.phys.latech.edu/$\sim$neven/ab/}
\bibitem{Bere82} Beresteckii V B, Lifshitz E M, and Pitaevskii L P 1982 
{\it Quantum Electrodynamics} (Oxford: Butterworth Heinemann).
\bibitem{Semon96} Semon M D and Taylor J R 1996 {\it Am. J. Phys.}  {\bf 64} 1361
\bibitem{Zhu90} Zhu X and Henneberger C 1990 {\it J. Phys. A: Math. Gen.} {\bf 23} 3983
\bibitem{Shel98} Shelenkov A L 1998 {\it Europhys. Lett.} {\bf 43} 623
\bibitem{Keat01} Keating J P and Robbins J M 2001 {\it J. Phys. A: Math. Gen.} {\bf 34} 807
\bibitem{Gronn06} Gronniger G, Simmons Z, Gilbert S, Caprez A, and Batelaan H 2006, (preprint)
\bibitem{Hors05} Horsley S A R and Babiker M 2005 {\it Phys. Rev. Lett.}{\bf 95} 010405
\bibitem{Hors07} Horsley S A R and Babiker M 2007 {\it J. Phys. B: At. Mol. Opt. Phys.} {\bf 40} 2003
\end{thebibliography}
\end{document}